\documentclass[sigconf]{acmart}
\usepackage[utf8]{inputenc}
\usepackage{amsthm}
\usepackage{amsmath}
\usepackage{bm}
\usepackage{soul}
\usepackage{mathtools}
\usepackage{thmtools,thm-restate}
\usepackage{enumitem}
\usepackage{pifont}
\usepackage{array}
\usepackage{float}
\usepackage{color}
\usepackage{caption}
\usepackage{subcaption}
\usepackage{multirow}
\usepackage{booktabs}
\usepackage{graphicx}
\usepackage{colortbl}
\usepackage{pythonhighlight}
\usepackage{hyperref}
\usepackage{comment}
\usepackage{mathtools}
\usepackage{todonotes}
\usepackage{hyperref}
\usepackage{wrapfig}

\usepackage{caption}
\usepackage{algorithm}
\usepackage[noend]{algpseudocode}
\include{marcos}

\DeclareMathOperator*{\argmin}{arg\,min}

\newcommand{\diag}{\mathbf{diag}}

\newcommand{\red}[1]{{\color{red}#1}}
\newcommand{\vol}{\operatorname{vol}}
\newcommand{\supp}{\operatorname{supp}}
\newcommand{\nei}{\operatorname{Nei}}
\newcommand{\mc}{\mathcal}
\newcommand{\eps}{\epsilon}

\newcommand{\R}{\mathbb R}

\newtheorem{theorem}{Theorem}
\newtheorem{lemma}[theorem]{Lemma} 
 
\newtheorem{remark}[theorem]{Remark}

\newcolumntype{P}[1]{>{\centering\arraybackslash}p{#1}}

\AtBeginDocument{%
  \providecommand\BibTeX{{%
    \normalfont B\kern-0.5em{\scshape i\kern-0.25em b}\kern-0.8em\TeX}}}
    
\setcopyright{acmcopyright}
\copyrightyear{2023}
\acmYear{2023}
\acmDOI{10.1145/1122445.1122456}

\acmConference[KDD '23] {Proceedings of the 29th ACM SIGKDD Conference on Knowledge Discovery and Data Mining}{August 6--10, 2023}{Long Beach, CA, USA.}
\acmBooktitle{Proceedings of the 29th ACM SIGKDD Conference on Knowledge Discovery and Data Mining (KDD '23), August 6--10, 2023, Long Beach, CA, USA}
\acmPrice{15.00}
\acmISBN{979-8-4007-0103-0/23/08}

\begin{document}

\title{Accelerating Personalized PageRank Vector Computation}

\author{Zhen Chen}
\email{zhenchen21@m.fudan.edu.cn}
\affiliation{%
  \institution{Fudan University, \\ Shanghai Key Laboratory of Data Science, Shanghai, China}
    \city{}
  \country{}
}

\author{Xingzhi Guo}
\email{xingzguo@cs.stonybrook.edu}
\affiliation{%
  \institution{Stony Brook University}
  \city{Stony Brook}
  \country{USA}
}

\author{Baojian Zhou}
\authornote{Corresponding Author.}
\email{bjzhou@fudan.edu.cn}
\affiliation{%
  \institution{Fudan University,\\ Shanghai Key Laboratory of Data Science, Shanghai, China}
      \city{}
  \country{}
}

\author{Deqing Yang}
\email{yangdeqing@fudan.edu.cn}
\affiliation{%
  \institution{Fudan University,\\ Shanghai Key Laboratory of Data Science, Shanghai, China}
    \city{}
  \country{}
}

\author{Steven Skiena}
\email{skiena@cs.stonybrook.edu}
\affiliation{%
  \institution{Stony Brook University}
  \city{Stony Brook}
  \country{USA}
}

\begin{abstract}
Personalized PageRank Vectors are widely used as fundamental graph-learning tools for learning graph embeddings, training graph neural networks, and detecting anomalous spammers. The well-known \textit{local} \textsc{FwdPush} algorithm \cite{andersen2006local} approximates PPVs and has a sublinear rate of $\mc{O}\big(\frac{1}{\alpha\eps}\big)$. A recent study \cite{wu2021unifying} found that when high precision is required, \textsc{FwdPush} is similar to the power iteration method, and its run time is pessimistically bounded by $\mc{O}\big(\frac{m}{\alpha} \log\frac{1}{\eps}\big)$. This paper looks closely at calculating PPVs for both directed and undirected graphs. By leveraging the \textit{linear invariant property}, we show that \textsc{FwdPush} is a variant of Gauss-Seidel and propose a Successive Over-Relaxation based method, \textsc{FwdPushSOR} to speed it up by slightly modifying \textsc{FwdPush}. Additionally, we prove \textsc{FwdPush} has \textit{local linear convergence} rate $\mc{O}\big(\tfrac{\vol(\mc S)}{\alpha} \log\tfrac{1}{\eps}\big)$ enjoying advantages of two existing bounds. 
We also design a new local heuristic push method that reduces the number of operations by 10-50 percent compared to \textsc{FwdPush}. For undirected graphs, we propose two momentum-based acceleration methods that can be expressed as one-line updates and speed up nonacceleration methods by $\mathcal{O}(1/\sqrt{\alpha})$. Our experiments on six real-world graph datasets confirm the efficiency of \textsc{FwdPushSOR} and the acceleration methods for directed and undirected graphs, respectively.
\end{abstract}

\keywords{Personalized PageRank, large-scale graph, local linear convergence, Successive Over-Relaxation}

\maketitle

\section{Introduction}
\label{sect:introduction}
As fundamental graph-learning tools, Personalized PageRank Vectors (PPVs) \cite{jeh2003scaling} have been widely used in classic graph mining tasks such as detecting anomalous spammers \cite{benczur2005spamrank,andersen2007local,andersen2008robust,spirin2012survey}, and modern graph representation learning methods such as graph embeddings \cite{tsitsulin2018verse,postuavaru2020instantembedding,guo2021subset} and graph neural networks \cite{bojchevski2019pagerank,bojchevski2020scaling,gasteiger2019predict,klicpera2021directional,epasto2022differentially,gasteiger2019diffusion,hassani2020contrastive,chen2020simple,tong2021directed,tong2020digraph,zhang2018link}. PPVs effectively capture the local proximity of graph nodes, making them useful for training improved graph neural network models and designing effective clustering algorithms \cite{andersen2006local}. As a result, efficient computation of PPVs is crucial for the current field of graph representation learning. 

The well-known \textsc{FwdPush} algorithm \cite{andersen2006local, berkhin2006bookmark} is a widely used tool for computing PPVs due to its effectiveness in approximating PPVs, easy implementation, and local nature. The cost of each iteration of \textsc{FwdPush} is dependent only on the volumes of nodes near the target node, and its total run time complexity can be bounded by $\mathcal{O}\big(\frac{1}{\alpha\epsilon}\big)$, where $\alpha$ is the damping factor, and precision parameter $\epsilon$ controls the precision of the per-entry of PPV. This time complexity bound is independent of the graph structure, making \textsc{FwdPush} a preferred method over the power iteration method, which requires access to the entire graph in each iteration. However, a recent study \cite{wu2021unifying} showed that when high precision is required, \textsc{FwdPush} behaves more like the power iteration method, with a pessimistically bounded run time complexity of $\mathcal{O}\big(\frac{m}{\alpha}\log \frac{1}{\epsilon}\big)$, where $m$ is the number of edges in the graph. This bound only holds for $\epsilon < (2m)^{-1}$, and it is unclear whether there exists a logarithmic factor bound for $\epsilon \geq (2m)^{-1}$.

A natural question we address in this paper is \textbf{Q1.} whether \textsc{FwdPush} has a \textit{locally linear convergence rate}, meaning that the per-epoch complexity is locally dependent on the graph, but the total number of epochs is still bounded by a logarithmic factor $\mathcal{O}\big(\log \frac{1}{\epsilon}\big)$. To answer this question, a key observation is that when $\alpha$ is close to 1, the local push method will not explore nodes far from the current target node, and thus the total run time per epoch remains local. It has been proven that when the graph is undirected, \textsc{FwdPush} is essentially a coordinate descent method, and computing an approximate PPV corresponds to an $\ell_1$-regularized optimization problem, which has a sparse solution \cite{fountoulakis2019variational}. This suggests that \textsc{FwdPush} is truly local. However, this equivalence is based on the assumption that the graph is undirected and is still unknown \textbf{Q2.} whether a similar optimization algorithm equivalence exists for directed graphs.

Questions Q1 and Q2 motivate us to study the PPV computation further. 
In this paper, we show for the first time that the well-known \textsc{FwdPush} algorithm is a variant of Gauss-Seidel when the underlying graph is directed. This is due to the \textit{linear invariant property} of \textsc{FwdPush}, which means that the updates of Gauss-Seidel for each coordinate of the target linear system are equal to the residual updates of FwdPush. We then propose to use the Successive Over Relaxation (SOR) based method to speed up \textsc{FwdPush}, namely \textsc{FwdPushSOR}.  The advantage of \textsc{FwdPushSOR} is that it speeds up the original method, and it can be naturally applied to other variants of FwdPush \cite{wu2021unifying}, even ones for dynamic graph settings \cite{zhang2016approximate}. Furthermore, to study the convergence rate better, we prove a \textit{locally linear convergence rate} $\mc{O}\big(\tfrac{\vol(\mc S)}{\alpha} \log\tfrac{1}{\eps}\big)$ where the precision is any positive number $\eps >0$ and $\vol(\mathcal{S})$ is the expected volume of a subset of \textit{active nodes} explored. Our analysis is simplified by adding a dummy node to the queue and considering only the non-zero residual nodes. 
For undirected graphs with small $\alpha$, momentum acceleration-based methods can accelerate by a factor of O($1/\sqrt{\alpha}$). Both acceleration-based methods can be implemented in one-line iteration updates. Our contributions are summarized as follows
\begin{itemize}
\item For the first time, we show that the well-known \textsc{FwdPush} algorithm is a variant of Gauss-Seidel. To further improve its performance, we propose \textsc{FwdPushSOR}, \textit{a speed-up local method} of the original \textsc{FwdPush} based on the SOR technique. \textsc{FwdPushSOR} is as simple as that of \textsc{FwdPush}.

\item We prove a \textit{locally linear convergence rate} of \textsc{FwdPush} with a complexity of $\mathcal{O}\big(\frac{\vol(\mathcal{S})}{\alpha} \log \frac{1}{\epsilon}\big)$, which combines the advantages of existing bounds where the expected volume of explored nodes, $\vol(\mathcal{S})$, is locally dependent on the underlying graph. Based on this insight, we design \textsc{FwdPushMean}, a new local push method variant that reduces the number of operations by 10-50\% for most target nodes.
 
\item For undirected graphs, we propose using the Heavy-Ball (HB) and Nesterov Acceleration Gradient (NAG) method to calculate high-precision PPR vectors, which are $\mathcal{O}(1/\sqrt{\alpha})$ times faster than the power iteration and FwdPush variants. Our methods can be implemented in just one line of code.
\item We conduct experiments on six real-world graphs and find that, FwdPushSOR and acceleration methods can significantly reduce run time and the number of operations required. For example, these local SOR methods are about 3 times faster on undirected graphs and about 2 times faster on directed graphs, respectively, when $\alpha = 0.15$.
\end{itemize}

The rest of the paper is organized as follows: In Sec. \ref{sect:related-work}, we discuss related work. Notations and the definition of PPV are presented in Sec. \ref{sect:preliminary}. Sec. \ref{sect:locality-fwdpush} gives the locality analysis of \textsc{FwdPush}. The accelerated algorithms for PPVs are presented in Sec. \ref{sect:momentum-ppr}. Finally, we present our experimental results and conclusions in Sec. \ref{sect:experiments} and Sec. \ref{sect:conclusion}, respectively. Our code and datasets are available at \textcolor{blue}{\url{https://github.com/ccczhen/AccPPR}}.

\section{Related Work}
\label{sect:related-work}

\textbf{PPVs and power iteration-based accelerations.\quad} 
The Personalized PageRank computation traces its roots to the work of \citet{jeh2003scaling}, who proposed using a personalized vector as a starting point for PageRank calculation instead of the uniform distribution used in the original PageRank algorithm \cite{page1999PageRank}. Acceleration methods for directed graphs are mainly based on power iteration, accessing the whole graph once per iteration, resulting in $\mc{O}(m)$ per-iteration updates. \citet{arasu2002pagerank} used Gauss-Seidel method to globally accelerate power iteration. \citet{kamvar2003extrapolation} proposed to use the Aitken extrapolation to speed up the PPVs calculation, but its effectiveness largely depends on the underlying graph and can sometimes lead to insignificant improvements. Other acceleration techniques, including the inner-outer loop approach \cite{gleich2010inner}, and methods proposed by \citet{lee2003fast} and systematically studied by \citet{langville2011google}, have also been proposed to improve the power iteration-based method. Different from these methods, this paper mainly focuses on accelerating local methods.

\textbf{\textsc{FwdPush} and its variants.\quad} The work of \citet{andersen2006local} proposed \textsc{FwdPush} (also known as an Approximate PageRank Approximation, APPR) algorithm for approximating PageRank personalization vectors. Later, it was used to approximate columns of the PPV matrix \cite{andersen2007local}. The algorithm has a sublinear time complexity bound of $\mathcal{O}(\tfrac{1}{\epsilon \alpha})$ due to significant residuals being pushed out from the residual vector per iteration. It is worth mentioning that the essentially same idea as the local push method is also proposed in \citet{berkhin2006bookmark}. Recent work by \citet{wu2021unifying} showed that for $\epsilon < (2m)^{-1}$, \textsc{FwdPush} converges more like power iteration methods, leading to a total run time complexity of $\mathcal{O}(\tfrac{m}{\alpha} \log \frac{1}{\epsilon})$. However, it remains unknown if there is an exponential improvement when $\epsilon > (2m)^{-1}$, which corresponds to local approximation. We take a step further in this direction.

A recent study by \citet{fountoulakis2019variational} found that computing PPV is equivalent to solving an $\ell_1$-regularized problem that can be treated as a variant of the coordinate descent algorithm. The reason is that the linear system can be reformulated as a quadratic strongly convex optimization problem when the graph is undirected. However, this analysis only works for undirected graphs, as the objective of the optimization requires a symmetric matrix. It is unclear how to apply the analysis to directed graph settings. When the graph is undirected, alternative global methods, such as the conjugate gradient method, exist but are complex to implement. Therefore, we explore using NAG and HB-based methods to speed up the computation. The question of whether there exists a locally independent bound $\mc{O}\big(\tfrac{1}{\sqrt{\alpha}\epsilon}\big)$ for the accelerated methods such as Accelerated Coordinate Descents \cite{lee2013efficient,allen2016even}, linear coupling \cite{allen2014linear} remains open asked by \citet{fountoulakis2022open}.

\textbf{Other related methods.\quad} Many Personalized PageRank-related algorithms \cite{lofgren2016personalized,jeh2003scaling,haveliwala2003topic,gleich2015pagerank,wang2020personalized} have been proposed, including for dynamic graph settings \cite{bahmani2010fast,zhang2016approximate}. These works mainly focus on computing the PPR for a single or subset of entries. The generalized PageRank problem has been reviewed in \cite{gleich2015pagerank} and symmetrically studied \cite{langville2011google,lofgren2015efficient}. One can find more details in \cite{gleich2015pagerank,langville2011google,lofgren2015efficient} and references therein. Our technique may be of independent interest to these directions.

\section{Preliminary}
\label{sect:preliminary}

\textbf{Notations.\quad} We consider unweighted simple graph $\mc{G} = (\mc{V},\mc{E})$ where $\mc{V} = \{1,2,\ldots, n\}$ is a set of nodes and $\mc{E} \subseteq \mc{V} \times \mc{V}$ is a set of edges with $m = |\mc{E}|$. The underlying graph $\mc G$ is either a directed graph or an undirected graph depending on the context. Bold lower letters are column vectors, e.g., $\bm p \in \R^n$. Bold capital letters, e.g., $\bm A \in \R^{n\times n}$ are matrices. $\bm D=\diag(d_1,d_2,\dots,d_n)$ is the diagonal out-degree matrix of $\mc{G}$.\footnote{See Appendix \ref{app:dangling-nodes} for dealing with dangling nodes.} The minimal and maximal degree is denoted as $d_{\min}$ and $d_{\max}$. $\nei(u)$ is the set of neighbors of $u$. $\bm A$ is denoted as the adjacency matrix of $\mc{G}$. The column stochastic matrix associated with $\bm A$ is defined as $\bm P:= \bm A^\top\bm D^{-1}$.\footnote{In case of $d_v = 0$ for some $v$, $\bm D^{-1}=\bm D^+$ where $\bm D^+$ is Moore–Penrose inverse of $\bm D$.} The teleportation parameter (a.k.a \textit{dumping factor}) $\alpha \in (0, 1)$ (usually $\alpha \in (0.0, 0.5)$ in practice). A vector $\bm r$ at time $t$ is denoted as $\bm r^t := [r_1^t, r_2^t, \ldots, r_n^t]^\top$. The volume of $\mc{S} \subseteq \mc V$ is defined as $\vol(\mc S) \triangleq \sum_{v \in \mc S} d_v$. The support of $\bm r$ is the set of nonzero indices, i.e., $\supp(\bm r) =\{ v : r_v \ne 0, v \in \mc V\}$. For any matrix $\bm M$, we denote $m_{i j}$ as the element of $\bm M$ at $i$-th row and $j$-th column. An indicating vector $\bm e_v \in \{\bm e_1,\bm e_2,\ldots, \bm e_n\}$ has value 1 in $v$-th entry and 0 otherwise. Similarly, $(\bm a)_u$ is an indicator vector with $a_u$ at $u$-th column and 0 otherwise. $\bm M_{i,:}$ is the $i$-th row of $\bm M$.

\subsection{Personalized PageRank Vector}

Given an underlying graph $\mc{G}=(\mc{V},\mc{E})$, an initial indicating vector ${\bm e_s}$, and a teleportation parameter $\alpha \in (0,1)$, PPV of a target node $s$ is a probability vector $\bm x$ such that
\begin{equation}
\bm x = \alpha \bm e_v + (1 - \alpha) \bm A^\top \bm D^{-1}  \bm x, \label{equ:ppr}
\end{equation}
where we call $\bm x$ satisfying Equ. (\ref{equ:ppr}) a PPV. The above equation is essential to access the $v$-th column of a nonnegative matrix, that is
\begin{equation}
\bm x = \alpha \left(\bm I - (1-\alpha)\bm A^\top \bm D^{-1}\right)^{-1} \bm e_v.\nonumber
\end{equation}
The definition of PPV is a generalization of Google matrix computation where $\bm e_s = \bm 1 / n$ used \cite{page1999PageRank}. The standard method of solving the linear system (\ref{equ:ppr}) is the fixed-point iteration working as the following
\begin{equation}
\bm x^{t+1} = (1-\alpha) \bm A^\top \bm D^{-1} \bm x^t +  \alpha \bm e_v. \nonumber
\end{equation}
As shown in \cite{gleich2015multilinear}, if we use $\bm x^0 = 0$, one immediately obtains that $\|\bm x^t - \bm x^*\|_1 = (1-\alpha)^t$ where we denote $\bm x^*$ as the true solution of the PPV for the target node $v$. 
 However, the above iteration method needs to access the whole graph, thus resulting in $\mc{O}(m)$ run time complexity per iteration. Next, we introduce the local  \textsc{FwdPush} method and explain how it can obtain a good approximation of PPV by only exploring a small set of nodes.

\subsection{\textbf{FwdPush} algorithm}

An efficient first-in-first-out queue-based implementation of \textsc{FwdPush} is presented in Algo. \ref{algo:forward-push}. At a higher level, \textsc{FwdPush} iteratively accesses nodes and their neighbors and moves a distribution $\bm r$ to another distribution $\bm x$. At each iteration, each node $u$ in $\mc Q$ satisfies $r_u \geq \eps d_u$. We call such a node $u$ an \underline{\textit{active node}}. Similarly, if a node $v$ has a low residual, i.e., $r_v < \eps d_v$, then it is \underline{\textit{inactive}}. Initially, \textsc{FwdPush} has the source node $s$ in $\mc{Q}$. With each iteration, it maintains the updates of two vectors $\bm r$ and $\bm x$, where $\bm r$ is a \underline{\textit{residual vector}}, and $\bm x$ is the \underline{\textit{estimation vector}}. For each active node $u$, it pushes $\alpha$ magnitudes of $r_u$ to $\bm x$ and spreads $(1-\alpha) r_u$ uniformly to $\nei(u)$. \begin{wrapfigure}{L}{0.26\textwidth}
\vspace{-3mm}
    \begin{minipage}{0.26\textwidth}
      \begin{algorithm}[H]
\caption{\small $\textsc{FwdPush}(\mc{G},\epsilon,\alpha, s)$}
\begin{algorithmic}[1]
\small
\State Initialization: $\bm r = \bm e_s, \bm x = \bm 0$
\State $\mc{Q} = [s]$
\While{$\mc{Q} \ne \emptyset$}
\State $u= \mc{Q}\text{.pop}()$
\State $x_u = x_u + \alpha \cdot r_u$
\For{$v \in \nei(u)$}
\State $r_v = r_v + \frac{(1-\alpha) r_u}{d_u}$
\If{$r_v \geq \eps d_v$ \& $v \notin \mc Q$}
\State $\mc{Q}\text{.push}(v)$
\EndIf
\EndFor
\State $r_u = 0$
\EndWhile
\State \textbf{Return} $\bm x$
\end{algorithmic}
\label{algo:forward-push}
\end{algorithm}
    \end{minipage}
  \end{wrapfigure}
  
\noindent It terminates when all $r_u < \eps d_u$. During these push operations, one always have $r_u, x_u \geq 0$ and $\|\bm r\|_1 + \|\bm x\|_1 = 1$. It can be shown that returned $\bm x$ is guaranteed by $|x_u - x_u^*| \leq \epsilon d_u, \forall u \in \mc V$ (See details in \cite{andersen2006local}). The essential effectiveness of \textsc{FwdPush} is because entries $\bm x$ indexing by nodes near to $s$ have large magnitudes, and entries of $\bm x$ follow a power-law distribution as demonstrated in Fig. \ref{fig:power-law} in the Appendix. Therefore, \textsc{FwdPush} quickly approximate $\bm x$ by only exploring these nodes \textit{near} to it. We aim to improve the speed of this local method. In the following section, we will present our key findings.

\section{Locality analysis of \textsc{FwdPush}}
\label{sect:locality-fwdpush}

This section presents the equivalence between \textsc{FwdPush} and a variant of Gauss-Seidel (G-S) and introduces a faster method based on the SOR technique. We demonstrate that \textsc{FwdPush} has a \textit{locally linear convergence rate} and offer insights that could help find more effective variants.

\subsection{\textsc{FwdPush} is a variant of Gauss-Seidel}
To show that \textsc{FwdPush} is a variant of G-S iteration (See Section 10.1.1 of \cite{golub2013matrix}). Recall, for solving the linear system 
\begin{equation}
\bm M \bm x = \bm b, \label{equ:mx=b}
\end{equation}
the G-S iteration updates $\bm x$ using the following online iteration
\begin{equation}
x_u^{t+1} = \frac{1}{m_{u u}} \left( b_u - \sum_{j=1}^{u-1} m_{u j} x_j^{t+1} - \sum_{j=u+1}^n m_{u j} x_j^t \right), u \in \mc{S}_t,  \label{equ:gauss-seidel}
\end{equation}
where $t$ indexes the current super-iteration, $x_j^{t+1}$ are elements have been updated up to time $j^{t+1}$, and $x_j^t$ are entries will be updated. $\mathcal{S}^t$ presents the set of indices of $\bm x$ updated in $t$-th super-iteration. Note that we have $\mc S_t = \mc V$ for all $t$ in the standard G-S iteration. G-S iteration is usually for solving system \eqref{equ:mx=b} when $\bm M$ is \textit{strictly diagonally-dominant}. We note that $\bm I - (1-\alpha) \bm A^\top \bm D^{-1}$ is a strictly diagonally-dominant matrix for a simple graph. The following theorem presents the equivalence between the variant of the Gauss-Seidel iteration and the \textsc{FwdPush} algorithm.

\begin{theorem}[\textsc{FwdPush} is Gauss-Seidel]
Each iteration updates $\bm r$ and $\bm x$ in Algo. \ref{algo:forward-push} of the \textsc{FwdPush}$(\mc G, \eps,\alpha, s)$ algorithm is equivalent to an iteration of the Gauss-Seidel iteration as defined in \eqref{equ:gauss-seidel} when $\bm b = \alpha \bm e_s$ and $\bm M = \bm I - (1-\alpha)\bm A^\top \bm D^{-1}$. Furthermore, $\mc{S}_t$ corresponds to the set of active nodes processed in Algo. \ref{algo:forward-push} at $t$-th epoch. \footnote{We will define an epoch of \textsc{FwdPush} by adding a dummy node presented in the next subsection.}
\label{thm:fwdpush-is-gauss-seidel}
\end{theorem}
\begin{proof}
The key to our proof is to use the \ul{\textit{linear invariant property}}. We state it as follows: Let $\bm x^t$ and $\bm r^t$ be the estimation and residual vector of calling \textsc{FwdPush}$(\mc G,\eps,\alpha,s)$ at time $t$, then for all $t\geq 0$, we have the following linear invariant property
\begin{equation}
\alpha \bm r^t = \alpha \bm r^0 - \big(\bm I -(1-\alpha) \bm A^\top \bm D^{-1}\big) \bm x^t. \label{equ:linear-invariant}
\end{equation}
To verify Equ. \eqref{equ:linear-invariant}, note that it is trivially true at the initial time $t=0$ where $\bm x^0 =\bm 0$. For all $t\geq 1$ and any active node $u$, notice that \textsc{FwdPush} updates $\bm x^{t-1}$ and $\bm r^{t-1}$ as the following
\begin{align}
{\bm x}^{t} &= \bm x^{t-1} + \alpha r_u^{t-1} \cdot \bm e_u \label{equ:forward-push-x}\\
\bm r^{t} &= \bm r^{t-1} - r_u^{t-1}\cdot \bm e_u  + (1-\alpha )r_u^{t-1} \bm A^\top \bm D^{-1} \bm e_u, \label{equ:forward-push-r}
\end{align}
where Equ. \eqref{equ:forward-push-x} corresponds to Line 5 and Equ. \eqref{equ:forward-push-r} represents Line6-10 of Algo. \ref{algo:forward-push} with the initial setup $\bm x^0 = \bm 0, \bm r^0 = \bm e_s$. To simplify Equ. \eqref{equ:forward-push-r}, one can reformulate it as
\begin{align*}
\alpha r_u^{t-1} \bm e_{u}  &= \alpha\left(\bm I - (1-\alpha) \bm A^\top \bm D^{-1}\right)^{-1} (\bm r^{t-1} - \bm r^{t}).
\end{align*}
$\bm x^t$ is thus the sum of the left-hand side of the above over $t$, that is,
\begin{align*}
\bm x^{t} &= \alpha\sum_{i=1}^{t} r_{u}^{i-1} \bm e_u = \alpha \left(\bm I - (1-\alpha) \bm A^\top \bm D^{-1}\right)^{-1} \sum_{i=1}^{t}(\bm r^{i-1} - \bm r^{i}) \\
&= \alpha \left(\bm I - (1-\alpha) \bm A^\top \bm D^{-1}\right)^{-1} \big(\bm r^0 - \bm r^{t}\big).
\end{align*}
Move the above-inverted matrix to the left; we see the linear invariant property \eqref{equ:linear-invariant} is valid. Next, we show \textsc{FwdPush} is a variant type of G-S iteration defined in \eqref{equ:gauss-seidel}. Since we assume $\bm M = \bm I - (1-\alpha)\bm A^\top\bm D^{-1}$ and $\bm b = \alpha \bm e_s$, then $m_{u u} = 1$ for the simple graph. The G-S iteration of \eqref{equ:gauss-seidel} can be rewritten as 
\begin{align*}
x_u^{t+1} &= x_u^t + \Big( b_u - \sum_{j=1}^{n} m_{u j} x_j^{t} \Big), \quad // x_j^{t} = x_j^{t+1} \text{ for } j < u 
\end{align*}
where we can use $\bm x^t$ to represent the updated $\bm x$ up to time $i^t$. Hence, each  update can be represented as a vector form as the following
\begin{align*}
\bm x^{t+1} &= \bm x^t + \left( \bm b - \bm M_{u,:} \bm x^t \right)_u \\
&= \bm x^t + \big(\alpha \bm r^0 - \big(\bm I -(1-\alpha) \bm A^\top \bm D^{-1}\big) \bm x^t \big)_u  \\
&= \bm x^t + \alpha r_u^t \bm e_u,
\end{align*}
where the second equality is from the definition of  $\bm b$ and $\bm M$, and the last equality follows from the linear invariant property \eqref{equ:linear-invariant}.
\end{proof}

When $\alpha$ is small, $\bm M$ has a large condition number corresponding to slow convergence of \textsc{FwdPush} and G-S. Fortunately, Thm. \ref{thm:fwdpush-is-gauss-seidel} immediately tells us that to speed up \textsc{FwdPush}, the acceleration technique used for G-S can also be applied for \textsc{FwdPush}.  To speed up the G-S procedure, we propose to use the Successive Over-Relaxation (SOR) technique \cite{young1954iterative,hackbusch1994iterative}, a well-known method for accelerating G-S of solving diagonally-dominant matrix. To update $\bm x^t$, by using SOR, we have (note $m_{uu}=1$)
\begin{align*}
\bm x^{t+1} &= (1-\omega) \bm x^t + \omega \Big( b_u - \sum_{j=1}^{u-1} m_{u j} x_j^{t+1} - \sum_{j=u+1}^n m_{u j} x_j^t \Big) \cdot \bm e_u \\
&= (1-\omega)\bm x^{t} + \omega \bm x^t + \omega(\bm b - \bm M_{u,:} \bm x^t)_u \\
&= \bm x^{t} + \omega \alpha r_u^t \bm e_u, \nonumber    
\end{align*}
where the relaxation parameter  $\omega \in (0, 2)$. The relaxed method is simply different by $\omega$ times. The next key is maintaining the invariant property for $\bm x^t$ and $\bm r^t$. To maintain the linear invariant property, we apply Equ. \eqref{equ:forward-push-x} and \eqref{equ:forward-push-r} $\omega$ times of original magnitudes. Hence, the corresponding residual updates of \textsc{FwdPush} become
\begin{align}
{\bm x}^{t} &= \bm x^{t-1} + \textcolor{blue}{\omega}\alpha r_u^{t-1} \cdot \bm e_u \label{equ:forward-push-sor-x}\\
\bm r^{t} &= \bm r^{t-1} - \textcolor{blue}{\omega} r_u^{t-1}\cdot \bm e_u  + \textcolor{blue}{\omega} (1-\alpha )r_u^{t-1} \bm A^\top \bm D^{-1} \bm e_u, \label{equ:forward-push-sor-r}
\end{align}
where we relax the assumption that entries of $\bm r^t$ could be negative. 
\begin{wrapfigure}{L}{0.28\textwidth}
\vspace{-5mm}
    \begin{minipage}{0.28\textwidth}
      \begin{algorithm}[H]
\caption{\small $\textsc{FwdPushSOR}(\mc{G},\epsilon,\alpha,s, \textcolor{blue}{\omega})$}
\begin{algorithmic}[1]
\small
\State Initialization: $\bm r = \bm 1_s, \bm x = \bm 0$
\State $\mc{Q} = [s]$
\While{$\mc{Q} \ne \emptyset$}
\State $u= \mc{Q}\text{.pop}()$
\State $x_u = x_u + \textcolor{blue}{\omega} \cdot \alpha r_u$
\For{$v \in \nei(u)$}
\State $r_v = r_v + \frac{\textcolor{blue}{\omega} (1-\alpha) r_u}{d_u}$
\If{$\textcolor{blue}{|r_v|} \geq \eps d_v$ and $v \notin \mc Q$}
\State $\mc{Q}\text{.push}(v)$
\EndIf
\EndFor
\State $r_u = (\textcolor{blue}{1-\omega}) r_u $
\EndWhile
\State \textbf{Return} $\bm x$
\end{algorithmic}
\label{algo:forward-push-sor}
\end{algorithm}
    \end{minipage}
  \end{wrapfigure}
  
  \noindent This violation enables us to move more magnitudes of residuals at once to $x_u$, thereby speeding up the entire procedure. Thus, based on the over-relaxed Equ. \eqref{equ:forward-push-sor-x} and \eqref{equ:forward-push-sor-r}, we propose \textsc{FwdPushSOR} shown in Algo. \ref{algo:forward-push-sor}, which is simple to implement and requires only the relaxation parameter $\omega$. The key invariant property of \textsc{FwdPushSOR} is that $\omega r_u$ magnitudes are \textit{excessively removed} from $r_u$ and distributed to all its outer neighbors and estimate $x_u$. Furthermore, \underline{this SOR-based method is still local}. It is worth noting that this approach can generally be applied to other variants of \textsc{FwdPush} \cite{wu2021unifying,zhang2016approximate}. For instance, the \textsc{PwrPush} algorithm proposed in \cite{wu2021unifying} can easily be modified to incorporate the SOR technique. We call this method \textsc{PwrPushSOR}.

\textbf{Parameter choosing for $\omega$.\quad} 
First of all, \textsc{FwdPushSOR} exactly recovers \textsc{FwdPush} when $\omega = 1$. Note that $\bm M$ is symmetric and positive-definite for undirected graphs, and SOR would converge on any $\omega \in (0,2)$ (see Thm. 4.4.12 of \cite{hackbusch1994iterative}). The optimal $\omega$ is 
\begin{equation}
\omega = 1+\left({\tfrac {1-\alpha}{1+{\sqrt {1-(1-\alpha)^2}}}}\right)^2 . \label{equ:optimal-omega}
\end{equation}
For directed graphs, choosing $\omega$ is more difficult since the matrix $\bm M$ is not easy to characterize. However, to choose $\omega$, one can use the following heuristic way: $\omega$ starts from the optimal value; if it fails, we decrease $\omega$ by a constant step until it reaches $1$. In practice, $\omega$ can reach about $1.5$ when $\alpha = 0.15$, which is more than two times faster than existing \textsc{FwdPush}. Yet, the convergence of SOR-based methods remains an open problem for future research.

Unlike the standard G-S iteration, \textsc{FwdPush} updates a subset of nodes $\mc S_t$ of $\mathcal{V}$ at each epoch. In a recent work of
\citet{wu2021unifying},  it has been proved that the queue-based implementation of \textsc{FwdPush} is similar to the power iteration method, hence for obtaining the final solution $\bm x^t$ with precision $\eps <(2m)^{-1}$, it requires $\mc{O}(\frac{m}{\alpha} \log \tfrac{1}{\epsilon})$ operations. In the following, we improve the analysis and show that \textsc{FwdPush} is locally linear convergent to $\bm x^*$. 

\begin{figure}
\centering
\includegraphics[width=.43\textwidth]{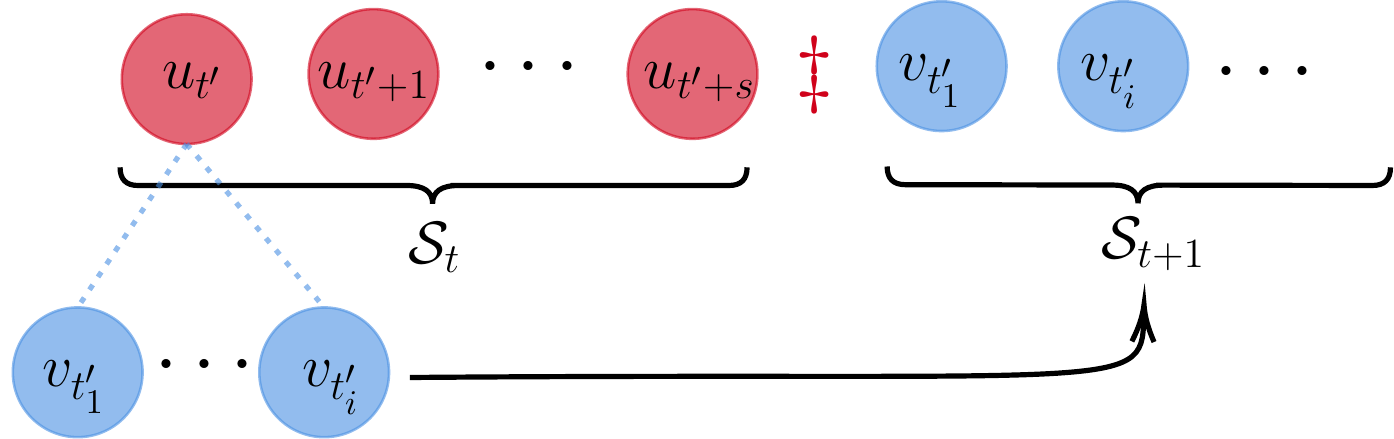}
\caption{The $t$-th epoch of \textsc{FwdPush} in Algo. \ref{algo:forward-push-dummy}. \textsc{FwdPush} maintains a queue $\mc Q$ which contains all active nodes. At the beginning of $t$-th epoch, $\mc S_t$ contains all  active nodes(red), which will be processed in $t$-th epoch. New active nodes (blue) generated in the current epoch will be processed in the next.\vspace{-4mm}}
\label{fig:t-super-epoch}
\end{figure}

\subsection{Local linear convergence of \textsc{FwdPush}}
To establish the \textit{locally linear convergence rate} and have a better illustration, we present slightly different \textsc{FwdPush} and add a \textit{dummy node} $\red{\ddag}$ as presented in Algo. \ref{algo:forward-push-dummy} where the dummy node helps to identify the super epochs. The parameter $t$ is indexing the epoch id, and $t'$ is indexing active node processing time, as shown in Fig. \ref{fig:t-super-epoch}.

The only difference between Algo. \ref{algo:forward-push-dummy} and Algo. \ref{algo:forward-push} is that Algo. \ref{algo:forward-push-dummy} always keeps $\red{\ddag}$ in the queue until no active nodes pushed into $\mc Q$. A new epoch begins whenever $\red{\ddag}$ pops out and is pushed into $\mc Q$ again. In the next theorem, we prove that \textsc{FwdPush} admits a \textit{local linear convergence} rate, and the total run time of \textsc{FwdPush} is only locally dependent on $\mc G$. We have the following notations: At the beginning of $t$-th epoch, the set of active and inactive nodes are denoted as $\mc S_t =\{u: r_u \geq \eps \cdot d_u, u \in \mc V\}$ and  $\mc U_t = \{v: 0< r_v < \eps \cdot d_v, v\in \mc V \}$, respectively. We also denote the support of $\bm r^t$ as $\mc I_t = \supp(\bm r^t)$. By these definitions, the total operations of \textsc{FwdPush} are the volume of all active nodes of all epochs, i.e., $\sum_{t=1}^T \vol(\mc S_t)$.

\begin{algorithm}[H]
\caption{$\textsc{FwdPush}(\mc{G},\epsilon,\alpha, s)$ with a dummy node}
\begin{algorithmic}[1]
\State Initialization: $\bm r = \bm e_s, \bm x = \bm 0$
\State $\mc{Q} = [s,\red{\ddag}]$ \quad\quad// Dummy node $\red{\ddag}$ at the end of $\mc Q$
\State $t = 0, t' = 0$
\While{$\mc{Q}\text{.size()} \ne 1$}
\State $u= \mc{Q}\text{.pop}()$
\If{$u ==\red{\ddag}$}
\State $\mc{Q}$\text{.push}$(u)$
\State $t = t + 1$ \quad\quad// Next epoch time
\State \textbf{continue}
\EndIf
\State $x_u = x_u + \alpha \cdot r_u$
\For{$v \in \nei(u)$}
\State $r_v = r_v + \frac{(1-\alpha) r_u}{d_u}$
\If{$r_v \geq \eps d_v$ and $v \notin Q$}
\State $\mc{Q}\text{.push}(v)$
\EndIf
\EndFor
\State $r_u = 0$
\State $t' = t' + 1$
\EndWhile
\State \textbf{Return} $\bm x$
\end{algorithmic}
\label{algo:forward-push-dummy}
\end{algorithm}

\begin{theorem}[Local linear convergence of FwdPush]
Let $\mc{S}_t$ and $\mc{U}_t$ be the set of active and inactive nodes, respectively.
Let $\bm x^t$ be the estimated PPR vector updated by \textsc{FwdPush} after $t$-th epoch, that is $ \bm x^t = \textsc{FwdPush}(\mc{G},\eps,\alpha,s)$. Then, for all $t \geq 0$, the $\ell_1$ estimation error $\|\bm x^{t+1} - \bm x^*\|_1$ has locally linear convergence rate, that is
\begin{equation}
\| \bm x^{t+1} - \bm x^*\|_1 \leq (1-\alpha \gamma_t)\|\bm x^{t}- \bm x^*\|_1, \label{inequ:local-convergence}
\end{equation}
where $\gamma_t$ is the local convergence factor $\gamma_t := \sum_{u \in \mc S_t} d_u  / \sum_{v \in \mc I_t} d_v$. Furthermore, the total run time of \textsc{FwdPush} is locally dependent on $\mc G$, is bounded by
\begin{equation}
\sum_{t=1}^T \vol(\mc{S}_t) \leq \frac{\vol(S_{1:T})}{\alpha\cdot\gamma_{1:T}}\log\left(\frac{C_{\alpha,T}}{\eps}\right), \label{inequ:operations-bound}
\end{equation}
where $\vol(S_{1:T})$ is the average volume, $\vol(S_{1:T}) = \sum_{t=1}^T\vol(S_t) / T$ and $\gamma_{1:T} = \sum_{t=1}^T\gamma_t / T$.
\label{thm:local-linear-convergence}
\end{theorem}

Before we prove the theorem, we introduce a key lemma, a new refinement from \citet{wu2021unifying} as the following.

\begin{lemma}[Locally linear decay of $\bm x^t$]
Let $\mc S_t$ and $\mc U_t $ be the set of \underline{\textit{active}} and \underline{\textit{inactive}} nodes at the beginning of $t$-th epoch with $t\in \{0,1,\ldots,t\}$, respectively. Then, after $t$-th epoch, we have $\| \bm x - \bm x^{t+1} \|_1 \leq \left( 1 - \alpha \gamma_t \right) \| \bm x - \bm x^{t}\|_1$, where $\gamma_t$ is the local convergence factor $\gamma_t := \sum_{u \in \mc S_t} d_u  / \sum_{v \in \mc I_t} d_v$.
\label{lemma:local-convergence}
\end{lemma}
\begin{proof}
We prove this lemma by showing that a significant residual has been pushed out from $\bm r^{t}$ to $\bm r^{t + 1}$ and corresponding gain from $\bm x^t$ into $\bm x^{t+1}$. The set of active nodes $\mc S_t$ has been processed: we use time $t'$ to index these nodes. The updates are the following.
For each $i$-th active node $u_{t_i'}$, the updates are from Line 10 to Line 15 of Algo. \ref{algo:forward-push-dummy} give us the following iterations

\begin{align}
\bm x_t &= \bm x_{u_{t_0'}} \xrightarrow[]{u_{t_1'}} {\bm x}_{t_1'}  \xrightarrow[]{u_{t_2'}} {\bm x}_{t_2'} \quad \cdots \quad \xrightarrow[] {u_{t_{|\mc{S}_t|}'}} {\bm x}_{t_{|\mc{S}_t|}'} = \bm x_{t+1} \nonumber\\
\bm r_t &= \bm r_{u_{t_0'}} \xrightarrow[]{u_{t_1'}} {\bm r}_{ u_{t_1'} }  \xrightarrow[]{u_{t_2'}} {\bm r}_{u_{t_2'}} \quad \cdots \quad \xrightarrow[]{u_{t_{|\mc{S}_t|}'}} {\bm r}_{u_{{|\mc{S}_t|}}'} = \bm r_{t+1}. \nonumber
\end{align}

For $t$-th epoch, the total amount of residual that had been pushed out is $\alpha \sum_{u_{t'} \in \mc{S}_t} r_{u_{t'}}$ (Line 10). That is,
\begin{equation}
\|\bm r_t\|_1 - \|\bm r_{t+1}\|_1 \geq \alpha \sum_{ u_{t'} \in \mc{S}_t} r_{u_{t'}}. \label{inequ:48}
\end{equation}
By the definition of $\mc S_t$ and $\mc U_t$, we have
\begin{align*}
\forall u_{t'} \in \mc{S}_{t}, r_{u_{t'}} \geq \eps \cdot d_{u_{t'}}, \quad \quad \forall v \in \mc{U}_t, 0 < r_v < \eps \cdot d_v.
\end{align*}
Summation above inequalities over all active nodes $u_{t'}$ and inactive nodes $v$, we have
\begin{equation}
\frac{\sum_{u_{t'} \in \mc{S}_{t}} r_{u_{t'}} }{\sum_{u_{t'} \in \mc{S}_{t}} d_{u_{t'}} } \geq \eps > \frac{\sum_{v \in \mc{U}_t} r_v}{\sum_{v \in \mc{U}_t} d_v }, \nonumber
\end{equation}
which indicates
\begin{align}
\frac{\sum_{u_{t'} \in \mc{S}_t} r_{u_{t'}} }{\sum_{u_{t'} \in \mc{S}_t} d_{u_{t'}} } &> \frac{\sum_{u_{t'} \in \mc{S}_t} r_{u_{t'}} + \sum_{v \in \mc{U}_t} r_v}{\sum_{u_{t'} \in \mc{S}_t} d_{u_{t'}} + \sum_{v \in \mc{U}_t} d_v } \nonumber\\
&= \frac{\sum_{v \in  \mc{I}_t} r_v}{\sum_{v \in \mc{I}_t}  d_v}  = \frac{\|\bm r_{t}\|_1}{\sum_{v \in  \mc{I}_t  } d_v} \label{inequ:key-inequality},    
\end{align}
where the last equality is due to the fact that $\mc{I}_t$ indexes all nonzero entries of $\bm r_t$, i.e., $\|\bm r_{t}\|_1 = \sum_{v \in \mc{I}_t } r_t(v)$. On the other hand, the $t$-th iteration error of \textsc{FwdPush} as $\|\bm x^{t} - \bm x^*\|_1$. Clearly, when $t=0$, $\|\bm x^{t} - \bm x^*\|_1 = 1$. For $t\geq 0$, we have
\begin{align*}
\|\bm x^* - \bm x^{t+1}\|_1 &= \|\bm x^* - \bm x^t \|_1 - \alpha \sum_{u_{t'} \in S_t} r_{u_t'}^t \\
&= \bigg(1 - \frac{\alpha \sum_{u_{t'} \in S_t} r_{u_t'}^t}{\|\bm r^t\|_1}\bigg)\|\bm x^* - \bm x^t \|_1 \\
&\leq \bigg(1 - \frac{\alpha \sum_{u \in \mc S_t} d_u }{ \sum_{v \in \mc I_t} d_v }\bigg)\|\bm x^* - \bm x^t \|_1,
\end{align*}
where the last inequality follows from \eqref{inequ:key-inequality}.
\end{proof}
Our new local convergence factor is $(1-\alpha \eta_t)$ which is strictly great than $1- \alpha \sum_{u\in S_t} d_u /m $ from \citet{wu2021unifying}, meaning better convergence rate. The other key ingredient of our theorem is to estimate the total number of epochs needed. The observation is that the total amount of residuals left in $\bm r^t$ is still \textit{relatively significant}, so the residual in the last epoch $\|\bm r^T\|_1$ is lower bounded. We state the upper bound of $T$ as the following.

\begin{figure}
\centering   \includegraphics[width=.45\textwidth]{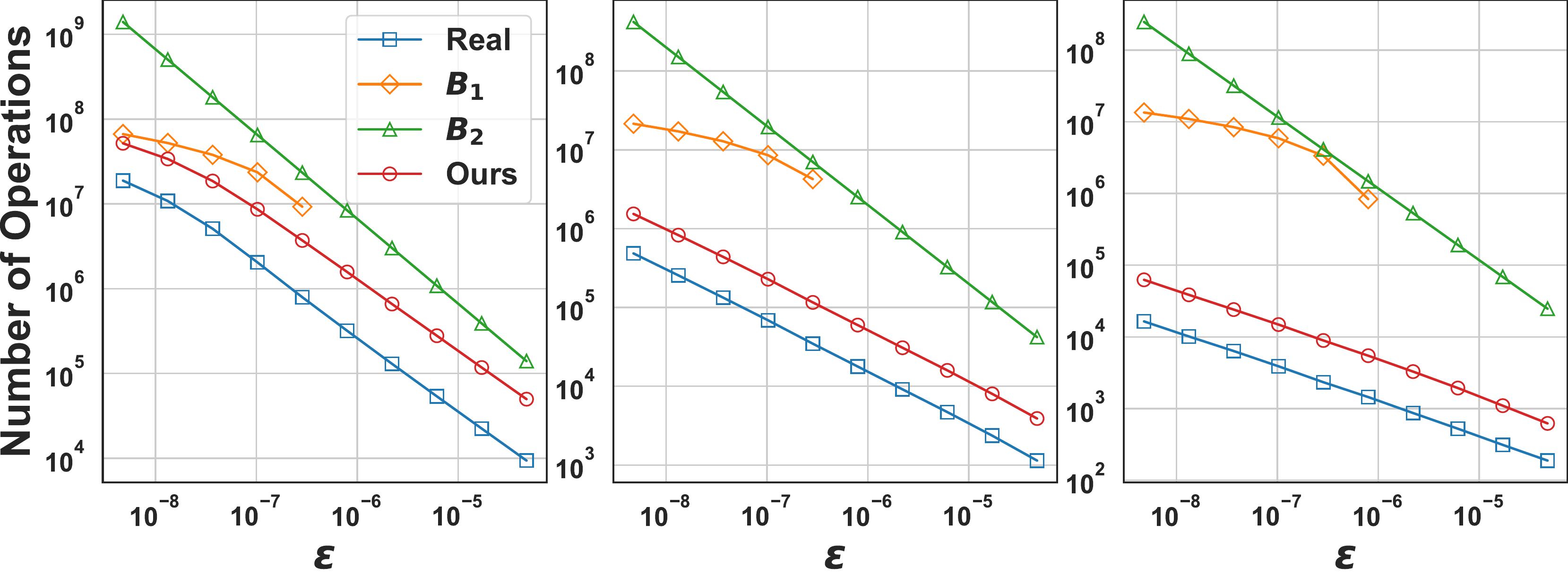}
    \caption{The number of operations estimated for the dblp dataset as a function of $\eps$. \textit{Real} stands for actual number of operations used in \textsc{FwdPush}, i.e. \textit{Real}$=\sum_{t=1}^T \vol(\bm S_t)$. $B_1 = \frac{m}{\alpha} \log(\frac{1}{\epsilon m}) + m$ provided in \cite{wu2021unifying}, $B_2=\frac{1}{\eps\alpha}$, and our new local bound. Left: $\alpha=0.15$, Middle: $\alpha=0.5$, and Right: $\alpha=0.85$. We randomly selected 100 nodes for each experiment and took the average of operations estimated. }
    \label{fig:ppr-upper-bounds}
\end{figure}

\begin{lemma}
Let $T$ be the total epochs used in \textsc{FwdPush}$(\mc G, \eps,\alpha, s)$, then it can be bounded by\begin{equation}
T \leq \frac{1}{\alpha\cdot\gamma_{1:T}} \log\left(\frac{C_{\alpha,T}}{\eps}\right),
\end{equation}
where $C_{\alpha,T} = 1/((1-\alpha)|\mc I_t|)$ and $\gamma_{1:T} = \sum_{t=1}^T \gamma_t/T$.
\label{lemma:total-epochs}
\end{lemma}
\begin{proof}
After the last epoch $T$, for each of nonzero node $v$, there was an active neighbor of $v$, denote as $u$, which pushed some residual $\frac{(1-\alpha)r_u}{d_u}$ to $v$. We denote each of this amount residual as $\Tilde{r}_v$, then for all $v\in \mc I_T$, we have
\begin{align*}
\|\bm r_T\|_1 = \sum_{v\in \mc I_T} r_v &\geq \sum_{v\in \mc I_T}\Tilde{r}_v := \sum_{v\in \mc I_T} \frac{(1-\alpha)r_u}{d_u} \\
&\geq \sum_{v\in \mc I_T} \frac{(1-\alpha) \eps d_u }{d_u} = \sum_{v\in \mc I_T} (1-\alpha) \eps = (1-\alpha)\eps|I_T|.    
\end{align*}
From \eqref{inequ:local-convergence} of Lemma \ref{lemma:local-convergence}, the upper bound of $\|\bm r_T\|_1$ is
\begin{equation*}
\| \bm r_T\|_1 \leq \prod_{t=1}^T \bigg(1 - \frac{\alpha \sum_{u \in \mc S_t} d_u }{ \sum_{v \in \mc I_T} d_v }\bigg) \| \bm x^* - \bm x^0\|_1 = \prod_{t=1}^T (1-\alpha \gamma_t).
\end{equation*}
Combine the lower and upper bound, we obtain
\begin{equation*}
(1-\alpha)\eps|I_T| \leq \prod_{t=1}^T (1-\alpha \gamma_t).
\end{equation*}
Take log on both sides of the above and use the fact $\log(1-\alpha \gamma_t) < -\alpha \gamma_t$, we reach
\begin{equation}
T \leq \frac{1}{\alpha\cdot \gamma_{1:T}} \log\left(\frac{C_{\alpha,T}}{\eps}\right), \nonumber
\end{equation}
where $C_{\alpha,T} = 1/((1-\alpha)|\mc I_T|)$ and $\gamma_{1:T} = \sum_{t=1}^T \gamma_t/T$.
\end{proof}
\begin{proof}[Proof of Theorem \ref{thm:local-linear-convergence}]
To obtain our main theory, we apply Lemma \ref{lemma:local-convergence} and Lemma \ref{lemma:total-epochs}, and notice that
\begin{align*}
\sum_{t=1}^T \vol(S_t) = T \cdot \vol(S_{1:T}) \leq \frac{\vol(S_{1:T})}{\alpha \cdot \gamma_{1:T}} \log\left( \frac{C_{\alpha,T}}{\eps} \right).
\end{align*}
\end{proof}
\begin{remark}
Our locality analysis provides intuition on the performance of local \textsc{FwdPush}. The total convergence rate is determined by $\gamma_t$, which is always a positive number. Note that $\gamma_0=1$ for the first epoch. The average volume accessed by \textsc{FwdPush} is certainly controlled by the following
\begin{equation}
\frac{\vol(\supp(\bm x_{\eps,\alpha}))}{T} \leq \vol(S_{1:T}) \leq \vol(\supp(\bm x_{\eps,\alpha})),
\end{equation}
where we simply denote $\bm x_T$ as $\bm x_{\eps,\alpha}$ which only depends on $\alpha$ and $\eps$ when $s$ and $\mc G$ are fixed. The above inequality indicates that when the solution is sparse, the volume will be much less than $m$. Another observation is that when $\eps \rightarrow 0$, we have $\gamma_{1:T} \rightarrow 1$, and $\vol(S_{1:T}) \rightarrow m$. Hence, it will recover to the power iteration-like method studied in \cite{wu2021unifying}. To see how the bound estimated the total operations, we conduct experiments on the dblp graph applying \textsc{FwdPush} over different $\eps$ as illustrated in Fig. \ref{fig:ppr-upper-bounds}. Compared to two known bounds, our parameterized local bound is tighter. We find a similar pattern on other graph datasets, detailed explained in the appendix.
\end{remark}

\textbf{Heuristic method \textsc{FwdPush-Mean}.\quad} Our parameterized bound in Equ. \eqref{inequ:operations-bound} involves a flexible quantity $\eta_T = \frac{\vol(S_{1:T})}{\gamma_{1:T}}$, which can guide us in finding better methods. For example, we can find a method that tries to minimize $\eta_T$. By noticing that $\eta_T$ is a trade-off between the ratio of active nodes processed in each epoch, we can either postpone pushing nodes with a large degree or the magnitudes $r_u$ are small. Doing this can save some operations and push more residuals for the next epoch. 

\begin{figure}[hbt!]
\centering
\includegraphics[width=.4\textwidth]{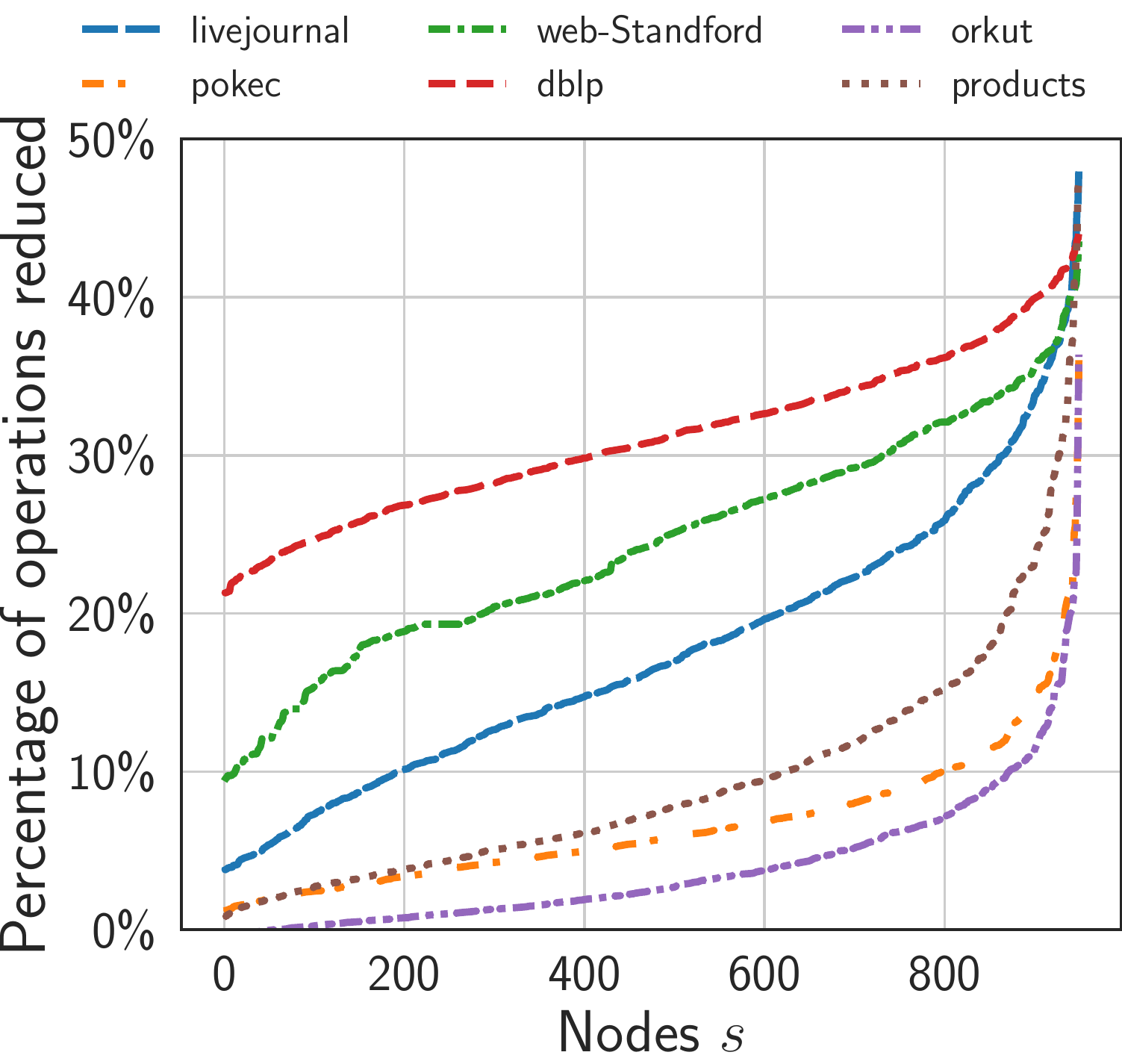}
\caption{The percentage of total operations reduced by \textsc{FwdPush-Mean} as a function of nodes over all six graphs. We fix $\eps=10^{-6}, \alpha=0.2$ and run both \textsc{FwdPush-Mean} and \textsc{FwdPush} on 1,000 randomly selected nodes from six graphs. The reduced percentage of operations is defined as the difference in the number of operations between two methods divided by the operations of \textsc{FwdPush}.}
    \label{fig:ppr-mean-operations}
\end{figure}

To make this idea concrete, at the beginning of each epoch, we push a subset of active nodes with a large magnitude ratio at each epoch, that is, to measure the \textit{magnitude factor} $r_u / d_u$. To do this, we check these nodes for each epoch by seeing whether the current magnitude factor of node $u$ is bigger than the mean of these ratios (which is an easy quantity to measure at the beginning of each super epoch). We postpone $u \in \mathcal{S}^t$ to the next epoch if its current magnitude factor satisfies
\begin{equation}
\frac{r_u}{d_u} < \bar{r} \triangleq \frac{1}{|\mc S_t|} \sum_{v\in S^t}\frac{r_v}{d_v}, \nonumber
\end{equation}
which means it is not worth pushing it; we can save $u$ to the next epoch so that $u$ accumulates more residual from $u$'s neighbors. We implemented this idea, called it \textsc{FwdPush-Mean}, and presented it in Algo. \ref{algo:forward-push-mean} of the appendix. Interestingly, \textsc{FwdPush-Mean} can effectively reduce the number of operations of \textsc{FwdPush}. Fig. \ref{fig:ppr-mean-operations} illustrates the number of operations reduced by the proposed \textsc{FwdPush-Mean} (see details of Algo. \ref{algo:forward-push-mean} in the appendix). Although run time may not be significantly reduced (observed in our experiments), this is still valuable in resource-limited scenarios as a significant amount of operations are reduced. 
 This method could also aid in finding a better heuristic for improving \textsc{FwdPush}. A detailed description of Algo. \ref{algo:forward-push-mean} can be found in the appendix.

Our local linear convergence guarantee for \textsc{FwdPush} allows for improvement through a trade-off between the number of active nodes explored in each epoch $\mathcal{S}^t$ and the total number of epochs $\mathcal{O}(\frac{1}{\alpha}\log \frac{C_{\alpha,T}}{\eps})$. If $\gamma_{1:T}$ is large, the total number of epochs is expected to be small, and $\log \frac{C_{\alpha,T}}{\eps}$ will also have a minimal effect. However, the total number of epochs will greatly depend on $\alpha$. If $\alpha$ is small, \textsc{FwdPush} becomes slow. Using SOR can speed up \textsc{FwdPush}. In the next section, we demonstrate that we can further save $1/\sqrt{\alpha}$ run time by employing global acceleration-based methods if the underlying graph is undirected.

\section{Momentum-based Methods for PPVs}
\label{sect:momentum-ppr}

This section analyzes the calculation of PPV when $\mc G$ is undirected. It reformulates the computation of PPV as a convex optimization problem and employs an acceleration-based technique to solve the linear system.

\subsection{Quadratic optimization lens}

We can rewrite the linear system of \eqref{equ:ppr} into strongly convex optimization and then transform the problem into a strongly convex optimization problem. Recall our target linear system is $(\bm I - (1-\alpha) \bm A \bm D^{-1} )\bm x = \alpha \bm e_s$. For an undirected graph, we multiply both sides by $\bm D^{-1/2}$ and rewrite the system such that the right-hand side matrix is symmetric; that is \eqref{equ:ppr} can be reformulated as the following
\begin{equation}
\big(\bm I - (1-\alpha) \bm D^{-1/2}\bm A \bm D^{-1/2} \big) \bm D^{-1/2} \bm x = \alpha \bm D^{-1/2} \bm e_s. \nonumber
\end{equation}
Denote $\tilde{\bm P} := \bm D^{-1/2}\bm A \bm D^{-1/2}$, ${ \bm y} := \bm D^{-1/2} \bm x$, and $\tilde{ \bm s} := \bm D^{-1/2} \bm e_s$. Notice that $\bm I - (1 - \alpha) \tilde{\bm P}$ is symmetric normalized Laplacian parameterized by $(1-\alpha)$. That is, we shall solve the linear system, $(\bm I - (1 - \alpha) \tilde{\bm P}) {\bm y} = \alpha \tilde{\bm s}$.  We define the minimization problem of a quadratic objective function $f$ as the following
\begin{equation}
\argmin_{\bm y \in \mathbb{R}^n} \left\{f(\bm y) \triangleq \frac{1}{2} {\bm y}^\top \Big( \bm I - (1 - \alpha) \tilde{\bm P} \Big) {\bm y} - \alpha \tilde{\bm s}^\top \bm y\right\},
\label{equ:quadratic-form}
\end{equation}
where $\tilde{\bm P}$ is not a row stochastic matrix any more but $\bm I - (1-\alpha) \tilde{\bm P}$ is still positive definite. Clearly, $f$ is a strongly convex function, by taking the gradient $\nabla f(\bm y)$ and letting it be zero, i.e. $\nabla f(\bm y) := (\bm I - (1 - \alpha) \tilde{\bm P}) {\bm y} - \alpha\tilde{\bm s} = \bm 0$, we see that $f$ has a unique solution $\bm y^* = \bm D^{-1/2} \bm x^*$. Therefore, the original solution $\bm x^*$ can be recovered from $\bm D^{1/2} \bm y^*$. To characterize the strongly convex and strong smoothness parameters of $f$ (See definitions of strongly convex and smoothness in Appendix \ref{app:section-b}), denote $\lambda_1, \lambda_2,\ldots, \lambda_n$ as eigenvalues of $\bm P$ with $\lambda_1=1 \geq \lambda_2 \geq \lambda_3 \geq \cdots \geq \lambda_n \geq -1$ and let $\tilde{\lambda}_1,\ldots,\tilde{\lambda}_n$ be eigenvalues of $\tilde{\bm P}$. Therefore, by the fact from the graph spectral theory \cite{chung1997spectral}, given the normalized $\bm I - \bm D^{-1/2} \bm A \bm D^{-1/2}$, we have its eigenvalues $0 = \tilde{\lambda}_1 \leq \cdots \leq \tilde{\lambda}_n \leq 2$. Therefore, the range of eigenvalues of  $\bm I - (1-\alpha) \tilde{\bm P}$ is $\lambda(\bm I - (1-\alpha) \tilde{\bm P}) \in [\alpha, 2 - \alpha]$.  

The above reformulation is commonly used in the optimization community and has also been studied in \citet{fountoulakis2019variational}. It was observed that when the graph is undirected,  \textsc{FwdPush} is a special case of a coordinate descent.   The coordinate descent for the above problem \eqref{equ:quadratic-form} is 
\begin{equation}
\bm y^{t+1} = \bm y^t - \eta_t \nabla_u f(\bm y^t) \cdot \bm e_u, \label{equ:coordinate-descent}
\end{equation}
where each step size should be chosen such that $\eta_t \leq 1/L_u$, where $L_u$ is the Lipschitz continuous parameter, $| \nabla_u f(\bm y + \delta \bm e_u) - \nabla_u f(\bm y)| \leq L_u \cdot \delta$, for all $\bm y \in \mathbb{R}^n$. Clearly, $L_u$ corresponds to the diagonal of $\bm I - (1-\alpha) \bm D^{-1/2} \bm A \bm D^{-1/2}$, which we always have $L_{u}\leq 1$. Replacing $\eta_t = 1$ and letting $-\bm \nabla(\bm y^t) = \bm r^t$ in \eqref{equ:coordinate-descent}. We can recover \textsc{FwdPush} accordingly. The coordinate descent algorithm could be accelerated by choosing momentum strategy \cite{allen2016even,lee2013efficient}.

The challenge of obtaining faster iteration based on the accelerated coordinate descent method remains open due to the lack of linear-invariant property for the momentum method \cite{fountoulakis2022open}. Nevertheless, we use standard momentum techniques and leverage the advantage of continuous memory. Our momentum-based method can be expressed in a single line, and local linear convergence suggests that the number of iterations is $\mathcal{O}\left(\frac{1}{\alpha \cdot\gamma_{1:T}} \log\frac{C_{\alpha,T}}{\eps}\right)$. In the next section, we will demonstrate that one can save $1/\sqrt{\alpha}$ run time.

\subsection{Accelerated methods}

\paragraph{Nesterov’s Accelerated Gradient (NAG) Method.} Given a strongly convex function $f$, NAG method \cite{nesterov2003introductory}, at $t$-th iteration, updates two vectors $\bm y^t$ and $\bm z^t$. It updates $\bm y^t$ with the help of $\bm z^t$ (with $\bm z^0 = \bm y^0$) as the following
\begin{align*}
\bm z^{t} &= \bm y^{t-1} - \eta_t \nabla f(\bm y^{t-1}) \\
\bm y^{t} &= \bm z^{t} + \beta_t (\bm z^t - \bm z^{t-1}).
\end{align*}
If we set $\eta_t = 1/(2 - \alpha)$ and $\beta_t = (1 - \kappa) / (1 + \kappa)$ with the inverse square root of condition number $\kappa = \sqrt{(2-\alpha) / \alpha}$, then we can write the iteration in one line as the following iteration procedure
\begin{equation}
\text{NAG:\quad} \bm y^{t+1} = 2 \left[\eta (\bm I + \tilde{\bm P}) \bm y^{t} \right] - (1-\kappa) \left[ \eta (\bm I + \tilde{\bm P}) \bm y^{t-1} \right] + \kappa^2 \tilde{\bm s}, \label{equ:nag-iteration}
\end{equation}
where $\eta = (1-\alpha) / ((2-\alpha)(1+\kappa))$ and $\bm y_0 = \bm 0, \bm y_1 = \alpha \bm D^{-1/2} \bm e_s$. Notice that $(\bm I + \tilde{\bm P}) \bm y^{t}$ can be used for the next iteration; hence the per-iteration operations are at most $m$. Then we have the following convergence rate for $\ell_1$ estimation error of $\bm x^*$. 

\begin{theorem}
Let $\bm y^{t+1}$ be the estimated vector returned by NAG method using iteration \eqref{equ:nag-iteration} (with $\bm z^0 = \bm y^0 = \bm 0$) and let $\bm x^t = \bm D^{1/2} \bm y^{t}$, then  the estimation error of $\bm x^*$ is upper bounded by
\begin{equation}
\| \bm x^{t} - \bm x^*\|_1 \leq d_{\max} \sqrt{\frac{2n}{\alpha}}  \exp \left(-\frac{t-1}{2\sqrt{(2-\alpha)/\alpha}}\right).
\end{equation}
And the total number of operations required for $\eps$-precision of per-entry of $\bm x^{t}-\bm x^*$, i.e., $\| \bm x^{t}-\bm x^*\|_\infty \leq \eps$ is
\begin{equation}
R_T = \mathcal{O}\left( \frac{m}{\sqrt{\alpha}} \log\left( \frac{d_{\max}}{\eps} \sqrt{\frac{2n}{\alpha}} \right)\right). \label{equ:better-bound}
\end{equation}
\label{thm:6}
\end{theorem}
\begin{proof}
The proof can be found at Appendix \ref{app:proof}.
\end{proof}
\begin{remark}
When high precision is required, and $\alpha$ is small, \textsc{FwdPush} is likely not a local method, meaning the per-epoch needs to touch the whole graph, hence the total complexity is about $\mc{O}(\tfrac{m}{\alpha} \log \frac{1}{\eps} )$. Compared with the bound provided in Thm. \ref{thm:nag-convergence}, NAG method is $1/\sqrt{\alpha}$ times faster. However, whether there exists a \textit{local} acceleration-based method like bound $\mathcal{O}\big(\frac{\vol(S)}{\sqrt{\alpha}} \log\tfrac{1}{\eps}\big)$ is still an open problem \cite{fountoulakis2022open}.
\end{remark}

\begin{figure*}[!ht]
    \centering
    \includegraphics[width=\textwidth]{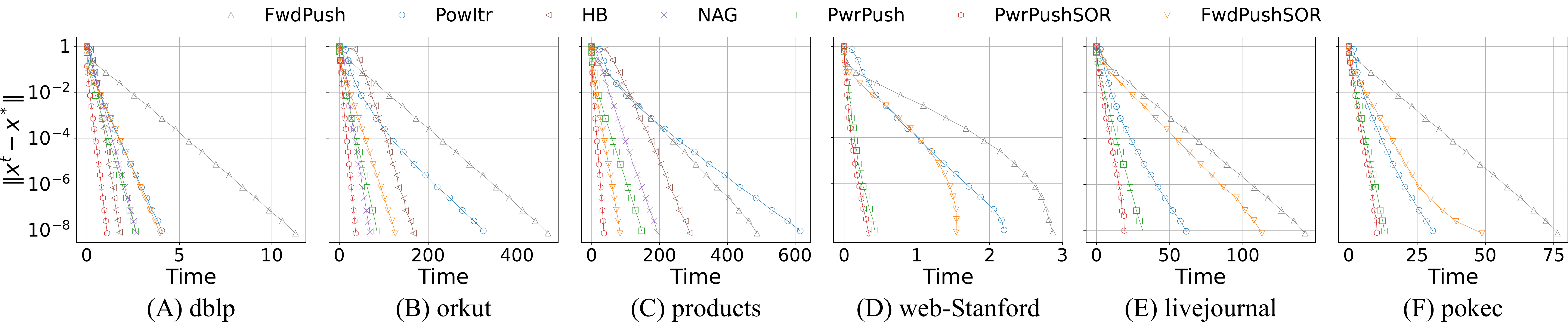}
    \caption{ Estimation error v.s. run time (seconds), $\alpha=0.15$.}
    \label{fig:time-0.15}
\end{figure*}

\begin{figure*}[!ht]
    \centering
    \includegraphics[width=\linewidth]{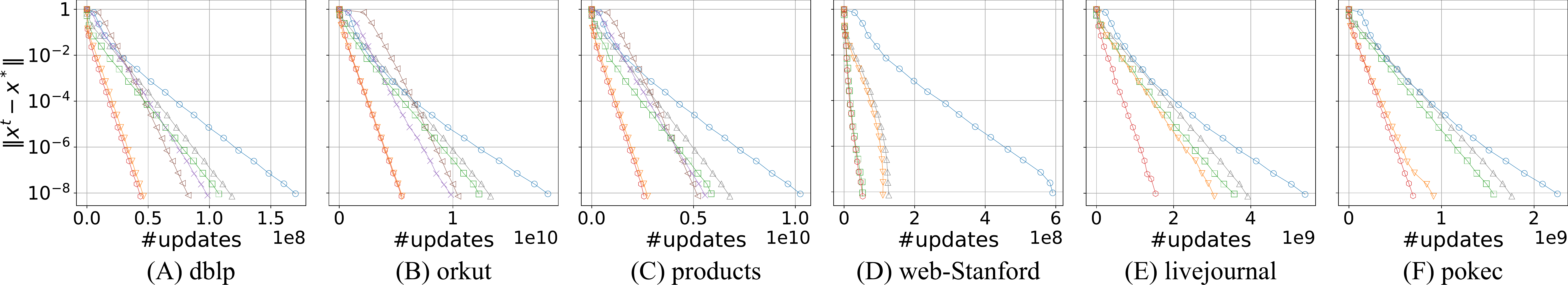}
    \caption{Estimation error v.s. \#residue updates (total operations), $\alpha=0.15$}
    \label{fig:oper-0.15}
\end{figure*}

\paragraph{Polyak's Heavy Ball (HB) method} Similar to the NAG method, the other popular momentum-based acceleration method is the Heavy Ball method \cite{polyak1964some}; different from the NAG method, the HB method updates $\bm y^t$ as the following
\begin{equation}
\bm y^{t+1} = \bm y^t - \eta_t \nabla f(\bm y^t) + \beta_t (\bm y^t - \bm y^{t-1}), \nonumber
\end{equation}
where we set $\eta_t = 4 / (\sqrt{2-\alpha} + \sqrt{\alpha})^2$ and $\beta_t = ((1-\kappa) / (1+\kappa))^2$, we can reach the following update method:
\begin{equation}
\bm y^{t+1} = 2 (1-\alpha) \frac{1+\kappa^2}{(1+\kappa)^2}\tilde{\bm P} \bm y^t - \frac{(1-\kappa)^2}{(1+\kappa)^2} \bm y^{t-1} + \frac{2\alpha(1+\kappa^2)}{(1+\kappa)^2} \tilde{\bm s}. \nonumber\\
\end{equation}
By changing variable $\bm D^{1/2} \bm y^{t+1} = \bm x^{t+1}$, we have HB 
\begin{equation}
\bm x^{t+1} =  \frac{2 \alpha(1+\kappa^2)}{(1+\kappa)^2}\bm A \bm D^{-1} \bm x^t - \frac{(1-\kappa)^2}{(1+\kappa)^2} \bm x^{t-1} + \frac{2(1-\alpha)(1+\kappa^2)}{(1+\kappa)^2} \bm v. \nonumber
\end{equation}

 Compared with local linear methods such as CD and \textsc{FwdPush}, the NAG and HB method admits a better convergence rate where the total number of iterations is about $\mathcal{O}(m/\sqrt{\alpha})$ compared with linear ones $\mc{O}(m/\alpha)$. Hence $\mc{O}(1/\sqrt{\alpha})$ times faster than methods presented in Sec. \ref{sect:locality-fwdpush}. One can also consider the most popular method, the conjugate gradient method \citep{shewchuk1994introduction}, which also has a better convergence rate than the standard gradient descent method and is the same as our momentum-based methods. However, our two methods are easier to implement. 

\section{Experiments}
\label{sect:experiments}

We conduct experiments on 6 real-world benchmark graphs to evaluate our proposed PPV algorithms, including \textsc{HB},  \textsc{NAG}, and \textsc{FwdPushSOR}, and \textsc{PwrPushSOR}, which is the combination of \textsc{PwrPush} \cite{wu2021unifying} and SOR. In the experiments, we aim to answer the following question: How fast are these proposed methods compared with baselines in terms of run time and the number of operations needed for different settings of $\alpha$? The results demonstrate the supreme efficiency boosted by SOR, achieving 2-3 times faster than strong baselines when $\alpha=0.15$. 

\noindent\textbf{Datasets.} 
We use both directed graphs (\textit{dblp} \cite{yang2012defining}, \textit{products} \cite{hu2020open} and \textit{orkut} \cite{yang2012defining})  as well as undirected graphs (\textit{web-Stanford} \cite{jung2017bepi}, \textit{pokec} \cite{takac2012data},\textit{livejournal} \cite{backstrom2006group}). These are the most common datasets for benchmark PPR algorithms.
We remove nodes with no in-degrees or out-degrees (dangling nodes), relabel the rest nodes, and use two directed edges to denote an undirected edge. 
Table \ref{tab:datasets} presents the detailed statistics.

\begin{center}
\vspace{-5mm}
\begin{table}[!h]
\caption{Datasets statistics}
\centering
\begin{tabular}{p{0.12\textwidth}p{0.08\textwidth}p{0.1\textwidth}p{0.09\textwidth}}
\toprule 
Dataset & $n$ & $m$ & Type of $\mc G$ \\ \midrule
dblp & 317,080 & 1,049,866 & undirected \\ 
products & 2,449,029 & 123,718,280 & undirected \\
orkut & 3,072,441 & 117,185,083 & undirected \\
web-Stanford & 281,903 & 2,312,497 & directed \\ 
livejournal & 4,847,571 & 68,993,773 & directed\\ 
pokec & 1,632,803 & 30,622,564 & directed \\\bottomrule
\end{tabular}
\vspace{-8mm}
\label{tab:datasets}
\end{table}
\end{center}

\noindent\textbf{Baselines.} 
We compare our proposed algorithms with three state-of-the-art baselines: \textsc{FwdPush} \cite{andersen2006local}, \textsc{PowItr}, and \textsc{PwrPush} \cite{wu2021unifying}, a variant of \textsc{FwdPush}. We use the proposed SOR technique to implement both \textsc{FwdPushSOR} and \textsc{PwrPushSOR}. For choosing the relaxation parameter $\omega$: 1) for undirected graphs, we directly use optimal value in \eqref{equ:optimal-omega}; 2) for directed graphs, we take an adaptive strategy described in Sec. \ref{sect:locality-fwdpush} to search $\omega$ where we set the minimal $\omega=1$ with step size 0.1 to the maximal value defined in Sec. \ref{sect:locality-fwdpush}. We only record the time consumed by the best $\omega$.  

\noindent\textbf{Experiment Settings.} 
For each graph, we uniformly sample 50 nodes for PPV calculation with  $\epsilon=\min\big( 1/10^8,1/m \big)$ (same as \cite{wu2021unifying}) and repeat experiments 5 times, recording the average running time, the number of residual updates, and the corresponding $\ell_1$-error. Note that the number of residual updates reflects the theoretical complexity regardless of the overheads incurred by various data structures. We measure PPV precision using $\ell_1$  error $\|\bm x^t - \bm x^*\|_1$ which can be measured by $\|\bm r^t\|_1$.

\noindent\textbf{Infrastructure and Implementation.} All experiments were conducted on a machine equipped with an Intel Xeon Gold 5218R CPU @ 2.10GHz (80 cores) with 256GB Memory. 
All algorithms are implemented in Python with the Numba library.\footnote{\url{https://numba.pydata.org/}}

\subsection{Experimental results}

\textbf{SOR-based methods are faster and need much less number of operations on both undirected and directed graphs.\quad} A-F of Fig. \ref{fig:time-0.15} and \ref{fig:oper-0.15} present the run time and the number of operations of PPV methods when $\alpha =0.15$, respectively. First, SOR-based methods are more than 2 times faster than their counterparts on both undirected and directed graphs shown in Fig. \ref{fig:time-0.15} (A-F). This confirms our SOR-based methods effectively speed up their counterparts. Note that by using a continuous memory access strategy, \textsc{PwrPush} and \textsc{PwrPushSOR} is, in general, faster than \textsc{FwdPush} and \textsc{FwdPushSOR} even if the number of operations for both is similar as shown in Fig. \ref{fig:oper-0.15} (A-F). Indeed, our SOR-based methods save half of the total operations. As shown in the appendix, we observed more significant speedup results when $\alpha=0.05$. The speedup advantages still exist even when $\alpha$ is large, i.e., $\alpha= 0.2$  and $0.25$. 

\textbf{Acceleration-based methods are faster and need fewer operations for undirected graphs.\quad} (A, B, C) of Fig. \ref{fig:time-0.15} and \ref{fig:oper-0.15} present results on undirected graphs. Among all methods, \textsc{HB} and \textsc{NAG} are faster than these non-acceleration methods but also use fewer operations. When $\alpha =0.05$ is a smaller value, the gap is more significant, as seen in Fig. \ref{fig:time-0.05} and \ref{fig:oper-0.05}. These results verify the $\mc{O}(1/\sqrt{\alpha})$ times faster predicted by our theorem. However, compared with these local methods, the speedup of acceleration-based methods is insignificant. One may expect a local version of acceleration-based methods could further improve \textsc{HB} and \textsc{NAG}. \textsc{PwrIter} as a global method uses more operations than \textsc{FwdPush} but requires less run time, as shown in the results of directed graphs. This is, again, because \textsc{PwrIter} uses a continuous memory access strategy while the nodes in the queue of \textsc{FwdPush} are randomly ordered, thus slowing down the process.

\begin{figure}
\centering
\includegraphics[width=.46\textwidth,height=2.8cm]{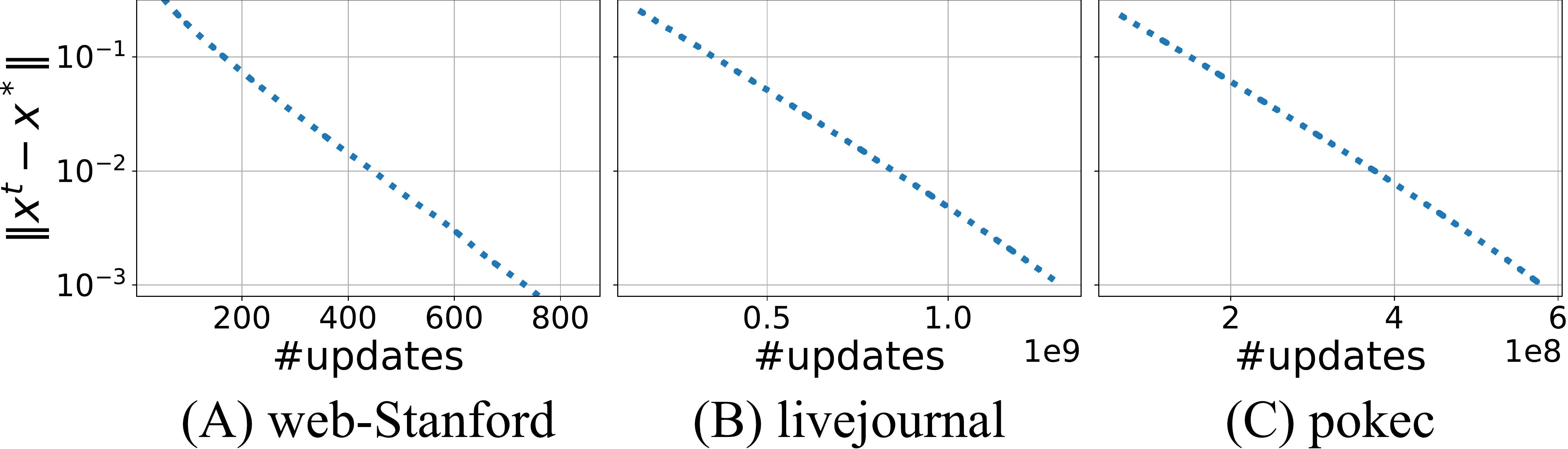}
\vspace{-2mm}
\caption{Locally linear convergence of \textsc{FwdPush}. For each directed graph, we randomly pick up a node and run \textsc{FwdPush} with $\eps=1/m$.\vspace{-4mm}} 
\label{fig:local-linear-convergence}
\end{figure}

\textbf{Local linear convergence rate of \textsc{FwdPush}.\quad} To empirically answer Q1 asked in Sec. \ref{sect:introduction}, we show that \textsc{FwdPush} has local linear convergence rate even when $\epsilon > (2m)^{-1}$. To do this, we simply set $\epsilon = 1/m$, randomly pick a node from three directed graphs, and then run \textsc{FwdPush}. The convergence rates are illustrated in Fig. \ref{fig:local-linear-convergence}. These linear decay rates of estimation error are consistent with Thm. \ref{thm:local-linear-convergence}. We found similar patterns on undirected graphs.

\section{Discussion}
\label{sect:conclusion}

This paper examines the calculation of PPV for directed and undirected graphs. We show that the commonly used local method for undirected graphs, \textsc{FwdPush}, is a variation of Gauss-Seidel. To improve the efficiency of \textsc{FwdPush}, we propose to use the SOR technique. Our SOR-based methods can successfully speed up current local methods significantly. Our SOR-based and acceleration methods could help build large-scale graph neural networks. It is worth seeing whether the SOR technique can be applied to the dynamic graph computation of PPVs. Additionally, we demonstrate that momentum-based acceleration methods can be used to obtain the PPV calculation for undirected graphs, providing $\mathcal{O}(1/\sqrt{\alpha})$ acceleration. Both acceleration methods are easy to implement and perform faster than other local methods when $\alpha$ is small. As a future work, it is interesting to see if it is possible to reduce $\mc{O}(m)$ from the bound in Thm. \ref{thm:nag-convergence} to a local quantity.

\section{acknowledgement}

The authors would like to thank the anonymous reviewers for their helpful comments. The work of Baojian Zhou is sponsored by Shanghai Pujiang Program (No. 22PJ1401300). The work of Deqing Yang is supported by Chinese NSF Major Research Plan No.92270121, Shanghai Science and Technology Innovation Action Plan No.21511100401. Steven Skiena was partially supported by NSF grants IIS-1926781, IIS-1927227, IIS-1546113, OAC-191952, and a New York State Empire Innovation grant.

\clearpage
\bibliographystyle{ACM-Reference-Format}
\bibliography{references.bib}

\clearpage
\onecolumn
\appendix

\section{Stochastic Matrix of $\mathcal{G}$ and dangling nodes}
\label{app:dangling-nodes}
Given the directed graph $\mathcal{G}$, we present several standard ways to construct row stochastic matrix $\bm P$. Recall $\bm D$ is the diagonal out-degree matrix of $\mathcal{G}$ and $\bm A$ is the associated adjacency matrix of $\mc{G}$. \textbf{Case 1.} If all nodes in $\mathcal{V}$ are not dangling nodes, that is, each node has at least one outgoing edge, then $\bm P = \bm D^{-1} \bm A$; \textbf{Case 2.} If some nodes are dangling nodes, two popular ways to create $\bm P$. Let $S = \{v: d_v = 0, v \in \mathcal{V}\}$ be the set of dangling nodes.
\begin{itemize}
\item For each dangling node, we create $n$ edges pointing all nodes, and the augmented degree matrix is $\bm D^\prime  = \bm D + n\operatorname{diag}(\bm 1_{S})$. So, $\bm P = \bm (D^\prime)^{-1} \bm A$.
\item We add a dummy node $v$ and create $|S|$ edges pointing from $S$ to $v$ meanwhile adding self-loop for node $v$. Hence, $\bm P = \left[\bm D^{-1}\bm A; [\bm 1_S^\top, 1]\right]$ where $;$ is the sign of row append.
\end{itemize}
For more options of creating $\bm P$, one can refer to Section 3 of \citet{gleich2015pagerank}.

\section{Strongly-convex and smooth of $f$ and the convergence of NAG method}
\label{app:section-b}
Given a convex function $f:\R^n \rightarrow \R$, we say $f$ is $\mu$-strongly convex if $\forall \bm x, \bm y\in \R^n$, we have
\begin{equation}
f(\bm x)-f(\bm y) \leq \bm \nabla f(\bm x)^{\top}(\bm x- \bm y)-\frac{\mu}{2}\|\bm x- \bm y\|_2^2. \nonumber
\end{equation}

We say $f$ is $L$-smooth, if for all $\bm x, \bm y \in R^n$, we have
\begin{equation}
 f(\bm x)-f(\bm y)- \bm \nabla f(\bm y)^{\top}(\bm x-\bm y) \leq \frac{L}{2}\|\bm x-\bm y\|_2^2. \nonumber
\end{equation}

Define Nesterov's Accelerated Gradient descent method (NAG) as the following iteration procedure
\begin{align*}
& \bm y_{t+1}= \bm x_t-\frac{1}{\beta} \bm \nabla f\left(\bm x_t\right), \\
& \bm x_{t+1}=\left(1+\frac{\sqrt{\kappa}-1}{\sqrt{\kappa}+1}\right) \bm y_{t+1}-\frac{\sqrt{\kappa}-1}{\sqrt{\kappa}+1} \bm y_t,
\end{align*}
where $\kappa = L / \mu$ is the condition number of $f$.

\begin{theorem}[\cite{bubeck2015convex}]
Let $f$ be $\mu$-strongly convex and $L$-smooth, then NAG method has the following convergence rate
\begin{equation}
f\left(\bm y_t\right)-f\left(\bm x^*\right) \leq \frac{\mu+L}{2}\left\|\bm x_1-\bm x^*\right\|_2^2 \exp \left(-\frac{t-1}{\sqrt{\kappa}}\right).
\end{equation}
\label{thm:nag-convergence}
\end{theorem}

\section{Proof of Theorem \ref{thm:6}}
\label{app:proof}
\begin{proof}
Notice that $f$ defined in \eqref{equ:quadratic-form} is $\alpha$-strongly convex and $(2-\alpha)$-strongly smooth. Hence letting $\mu = \alpha$ and $L = 2 - \alpha$,  by applying Thm. \ref{thm:nag-convergence}, we have 
\begin{align*}
f\left(\bm y^t\right)-f\left(\bm y^*\right) &\leq \frac{\mu+L}{2}\left\|\bm x^0 -\bm y^*\right\|_2^2 \exp \left(-\frac{t}{\sqrt{\kappa}}\right) \\
&= \left\|\bm x^0 -\bm y^*\right\|_2^2 \exp \left(-\frac{t}{\sqrt{(2-\alpha)/\alpha}}\right).    
\end{align*}
Note for any strongly convex function $f$, the optimization error can also be lower bounded by $\frac{\alpha}{2}\|\bm y^t - \bm y^*\|_2 \leq f(\bm y^t) - f(\bm y^*)$, then we reach
\begin{equation}
\frac{\alpha}{2} \| \bm y^t - \bm y^*\|_2^2 \leq \| \bm x^0 - \bm y^*\|_2^2  \exp \left(-\frac{t}{\sqrt{(2-\alpha)/\alpha}}\right), \nonumber
\end{equation}
as $ \|\bm x^0 - \bm y^*\|_2 = \|\bm y^*\|_2 \leq \|\bm y^*\|_1 = \| \bm D^{-1/2} \bm x^*\|_1 \leq \sqrt{1/d_{\min}} \leq 1$. Notice $\bm x^{t+1} = \bm D^{1/2} \bm y^{t+1}$, we then have
\begin{equation}
\|\bm D^{-1/2} (\bm x^{t+1} - \bm x^*)\|_2 \leq \sqrt{\frac{2}{\alpha}} \exp \left(-\frac{t}{2\sqrt{(2-\alpha)/\alpha}}\right) \nonumber
\end{equation}
Note for any $\bm x \in \R^n$, we have $\| \bm x\|_1 \leq \sqrt{n} \|\bm x\|_2$. Then we have
\begin{equation}
\|\bm x^{t+1} - \bm x^*\|_1 \leq d_{\max}\sqrt{\frac{2 n}{\alpha}} \exp \left(-\frac{t}{2\sqrt{(2-\alpha)/\alpha}}\right) \nonumber
\end{equation}
Use the fact $\|\bm x^{t+1} - \bm x^*\|_\infty \leq \|\bm x^{t+1} - \bm x^*\|_1$ and apply per-iteration operations $m$ of each iteration, we have the above operation complexity bound.
\end{proof}

\section{More experimental results}

\subsection{Power law distribution of PPVs}

\begin{figure}[H]
\centering
\includegraphics[width=1\linewidth]{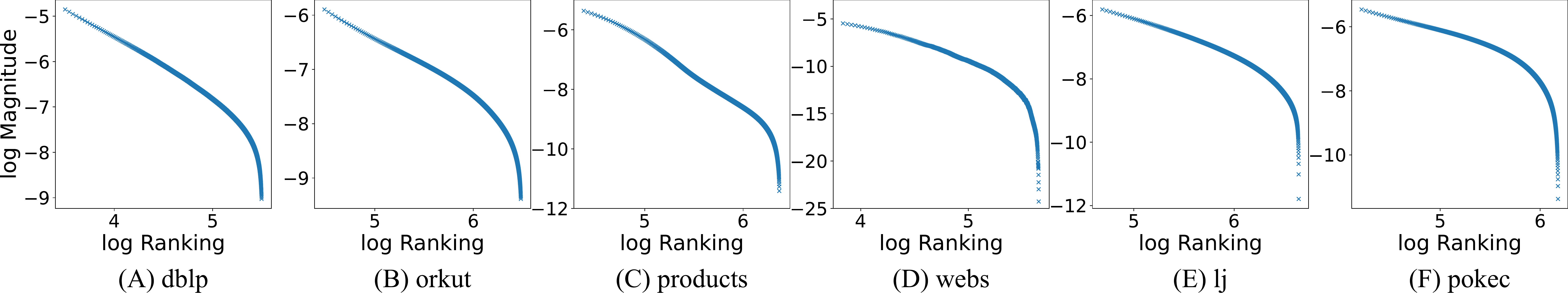}
\caption{The power law distribution of magnitudes $\bm x^*$ of six graphs. We randomly pick one node from a graph and run the power iteration algorithm to obtain high precision $\bm x^*$. It is important to note that in this particular setting, the power law distribution is characterized by a power law with a cutoff, as discussed in Clauset's work \cite{clauset2009power}.}
\label{fig:power-law}
\end{figure}

As shown in Fig. \ref{fig:power-law}, we present the magnitudes of $\bm x^*$ as a function of their ranks. We sort all magnitudes in descending order and label these magnitudes from rank 1 to rank $n$. Evidently, these magnitudes adhere to the power law distribution with a cutoff, as described by Clauset \cite{clauset2009power}. More specifically, let $p(x)$ represent a magnitude where $x$ is the associated ranking ID; this yields the following relation
\begin{equation}
p(x) \propto L(x) x^{-a}, \text{ where } L(x) = e^{-b x}. \label{equ:22}
\end{equation}
One can find suitable parameters $a$ and $b$ to fit these curves using \eqref{equ:22}.

\subsection{Comparison of empirical bounds}

\begin{figure}[H]
\centering
\includegraphics[width=1.\textwidth]{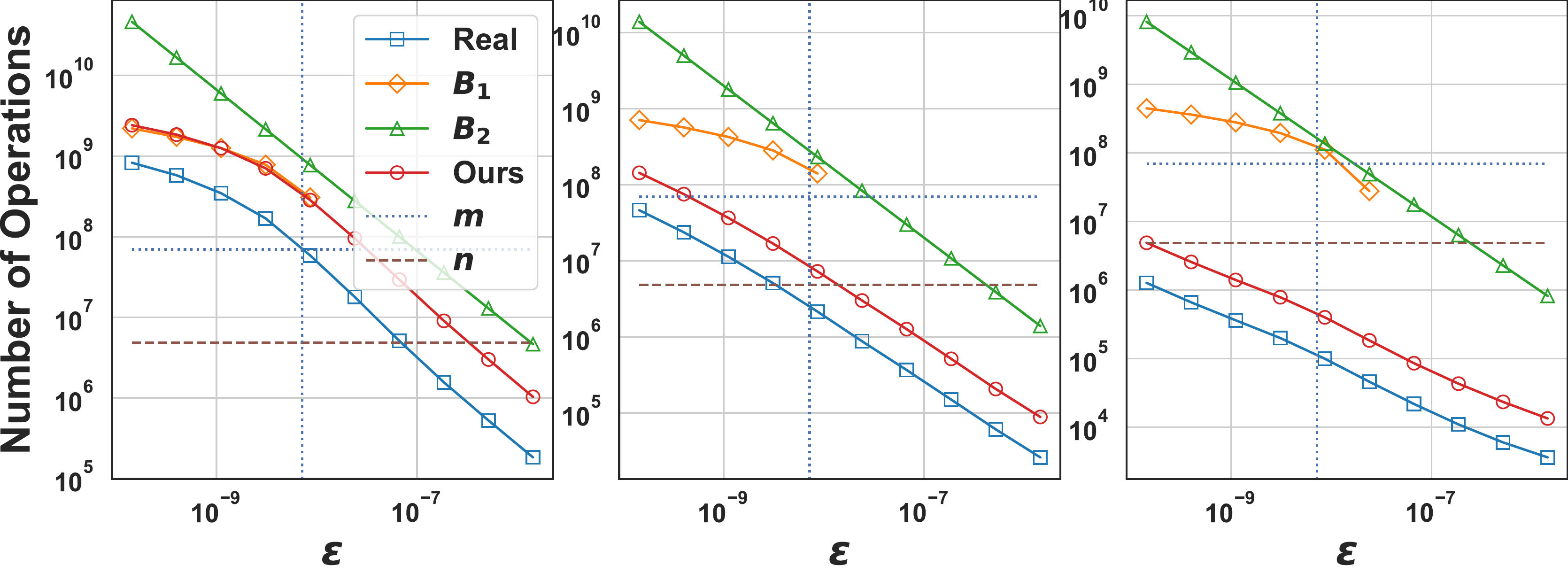}
\caption{The bounds of \textit{livejournal} dataset as a function of $\eps$. The vertical line is where $\epsilon=(2m)^{-1}$. Left: $\alpha=0.15$, Middle: $\alpha=0.5$, and Right: $\alpha=0.85$.}
\label{fig:livejournal-n-m}
\end{figure}

\begin{figure}[H]
\centering
\includegraphics[width=1.\textwidth]{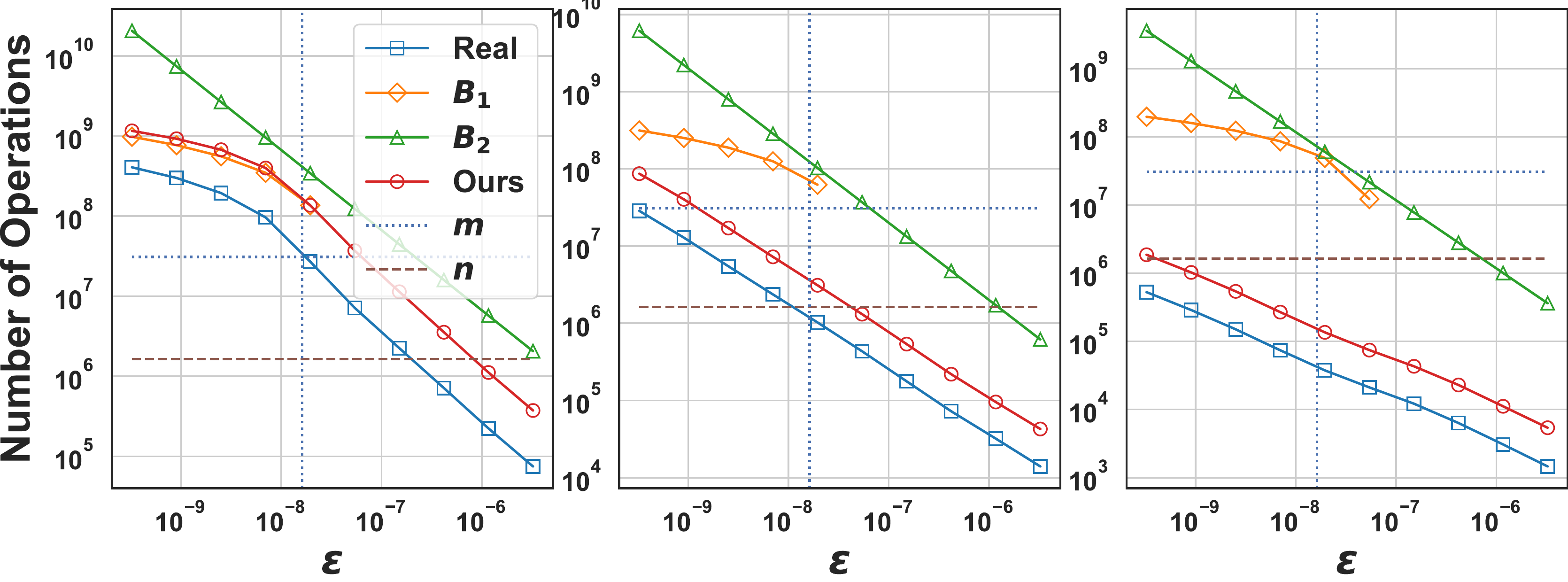}
\caption{The bounds of \textit{pokec} dataset as a function of $\eps$.}
    \label{fig:pokec-n-m}
\end{figure}

To further validate the effectiveness of our parameterized bound, we carry out a series of experiments on two additional graph datasets, specifically, \textit{livejournal} and \textit{pokec} shown in Fig. \ref{fig:livejournal-n-m} and \ref{fig:pokec-n-m}, respectively. When the value of $\alpha$ is relatively small, our bound is similar to $B_1$ for small values of $\eps$, and it is empirically tighter than $B_2$, irrespective of whether $\eps$ is small or large. As $\alpha$ increases, the comparative tightness of our bound becomes markedly more pronounced.

\subsection{More experiments on the run time and the number of operations comparison}

\begin{figure}[H]
    \centering
\includegraphics[width=\linewidth]{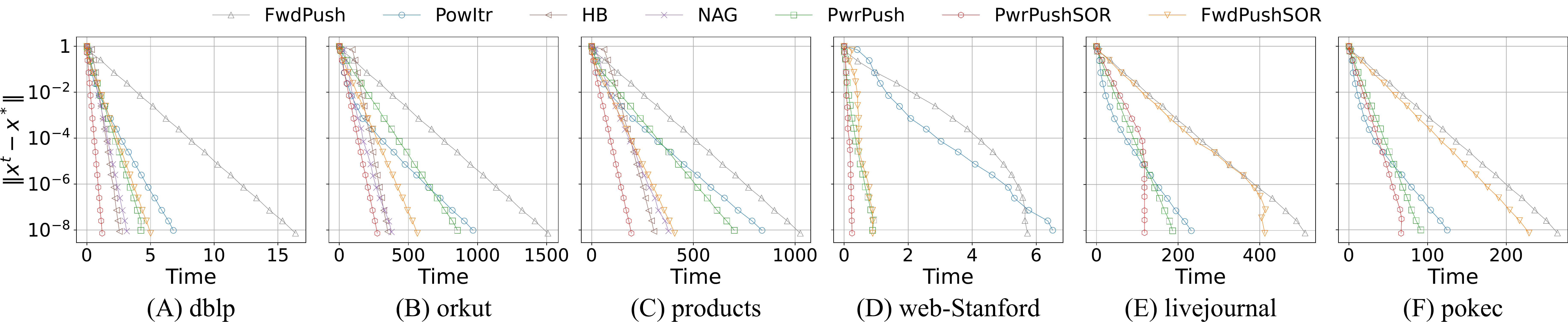}
    \caption{Actual $l_1$-error v.s. execution time (seconds), $\alpha=0.05$.}
    \label{fig:time-0.05}
\end{figure}

\begin{figure}[H] 
    \centering    \includegraphics[width=\linewidth]{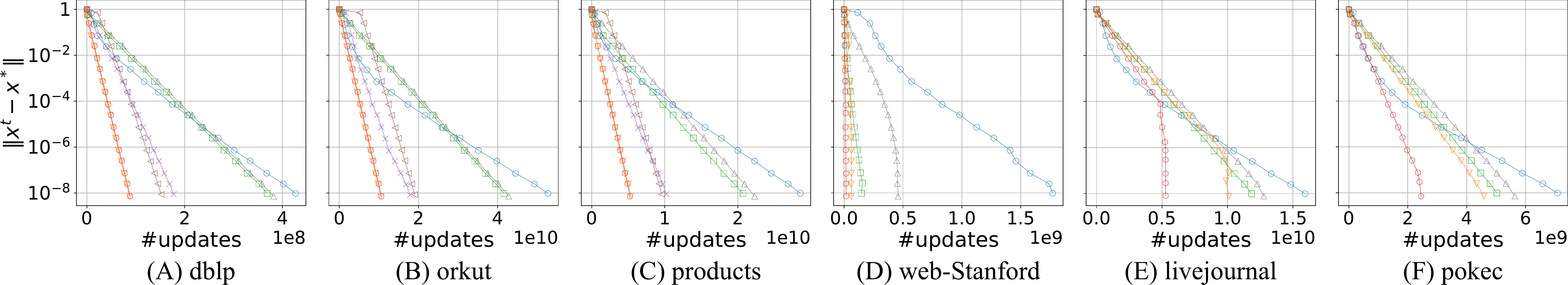}
    \caption{Actual $l_1$-error v.s. \#residue updates, $\alpha=0.05$}
    \label{fig:oper-0.05}
\end{figure}

\begin{figure}[H] 
    \centering
    \includegraphics[width=\linewidth]{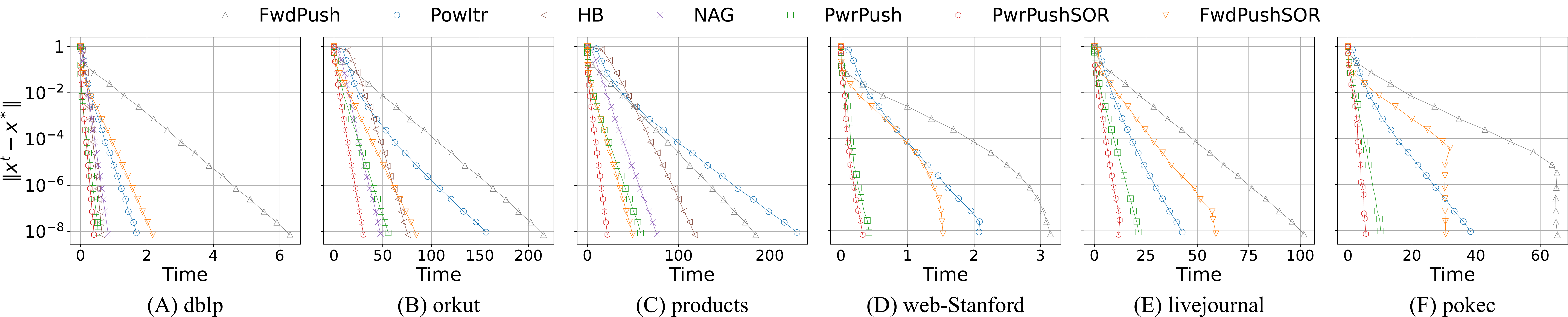}
    \caption{Actual $l_1$-error v.s. execution time (seconds), $\alpha=0.2$.}
    \label{fig:time-0.2}
\end{figure}

\begin{figure}[H]
    \centering
    \includegraphics[width=\linewidth]{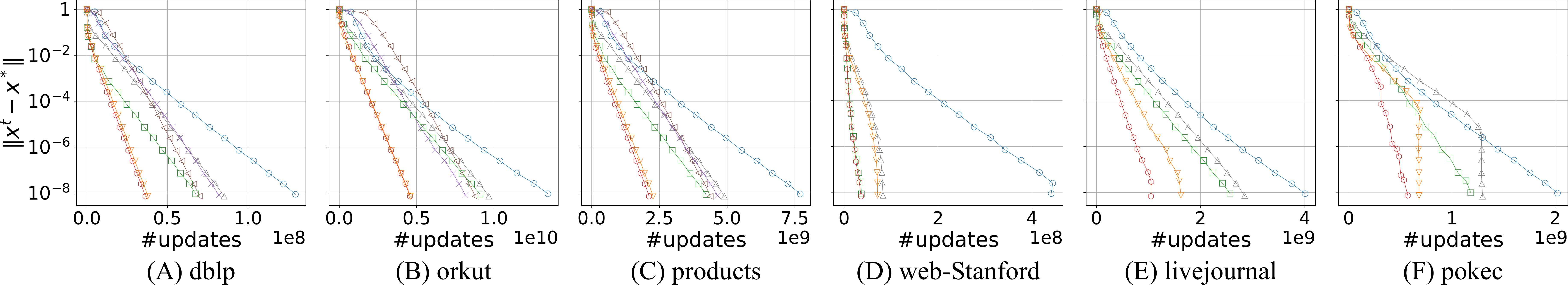}
    \caption{Actual $l_1$-error v.s. \#residue updates, $\alpha=0.2$}
    \label{fig:oper-0.2}
\end{figure}

A-F of Fig. \ref{fig:time-0.05}, \ref{fig:oper-0.05}, \ref{fig:time-0.2} and \ref{fig:oper-0.2} present the run time and the number of operations of PPV methods when $\alpha =0.05$ and $\alpha=0.2$, respectively. When compared to the setting where $\alpha = 0.15$, SOR-based methods exhibit a more significant improvement when $\alpha = 0.05$. For example, in the dataset of \textit{products}, our \textsc{PwrPushSOR} is more than 5 times faster than \textsc{FwdPush} method.

\textbf{Observation of superlinear behavior.\quad} During the course of our experiments, we discerned that both \textsc{PwrPushSOR} and \textsc{FwdPushSOR} exhibited the potential for superlinear behavior during the final few iterations. For example, when setting $\alpha$ at 0.05, the runtime required by \textsc{PwrPushSOR} displayed superlinearity in relation to $\ell_1$ error (see E of Fig. \ref{fig:time-0.05} and \ref{fig:oper-0.05}). Interestingly, this phenomenon mirrors the well-known superlinear convergence behavior observed when employing the conjugate gradient method to solve large symmetric systems of equations. This intriguing pattern certainly warrants further exploration and study.

\begin{figure}[H] 
    \centering
    \includegraphics[width=\linewidth]{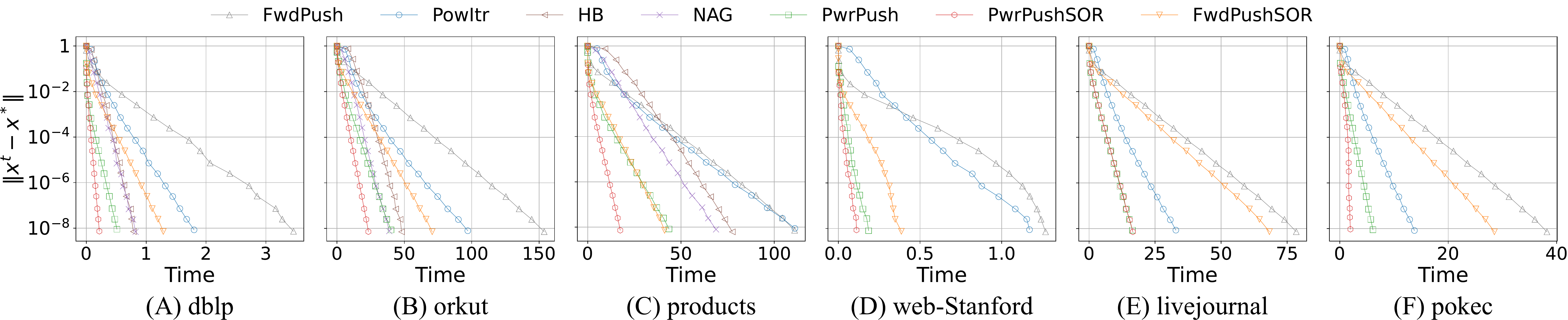}
    \caption{Actual $l_1$-error v.s. execution time (seconds), $\alpha=0.25$.}
    \label{fig:time-0.25}
\end{figure}

\begin{figure}[H] 
    \centering
    \includegraphics[width=\linewidth]{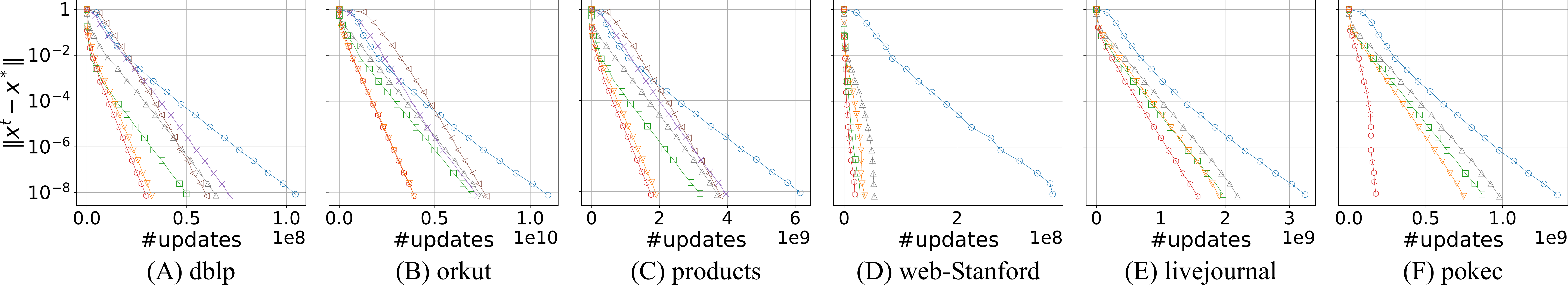}
    \caption{Actual $l_1$-error v.s. \#residue updates, $\alpha=0.25$}
    \label{fig:oper-0.25}
\end{figure}

\textbf{Significant speedup of local SOR methods even with large $\alpha$.\quad} As depicted in Fig. \ref{fig:time-0.25} and Fig. \ref{fig:oper-0.25}, it is evident that the local SOR method still requires fewer runtime operations to reach equivalent approximate solutions, even when $\alpha$ is large. This consistently efficient performance clearly underlines the effectiveness and versatility of our method across a broad spectrum of settings.

\clearpage
\section{\textsc{FwdPush-Mean} algorithm}

We present \textsc{FwdPush-Mean} in Algo. \ref{algo:forward-push-mean}. Compared with \textsc{FwdPush}, it computes the statistic of average residuals of active nodes $\bar{r}$ at Line 8. Residuals of nodes that are less than this average will postpone to the next epoch. Note the run time of computing $\bar{r}$ is not bigger than $|S_t|$; hence the total run time complexity will be the same as \textsc{FwdPush}. In practice, we found this strategy could help to reduce the total amount of push operations as shown in Fig. \ref{fig:ppr-mean-operations}.

\begin{algorithm} 
\caption{$\textsc{FwdPush-Mean}(\mc{G},\epsilon,\alpha, s)$ with a dummy node}
\begin{algorithmic}[1]
\State Initialization: $\bm r = \bm e_s, \bm x = \bm 0$
\State $\mc{Q} = [s,\red{\ddag}]$ \quad\quad// Dummy node $\red{\ddag}$ at the end of $\mc Q$
\State $t = 0, t' = 0$
\State $\bar{r} = 0$
\While{$\mc{Q}\text{.size()} \ne 1$}
\State $u= \mc{Q}\text{.pop}()$
\If{$u ==\red{\ddag}$}
\State $\bar{r} = \sum_{i\in {\mc S}^t} \frac{r_i / d_i}{|{\mc S}^t|} $
\State $\mc{Q}$\text{.push}$(u)$
\State $t = t + 1$ \quad\quad// Next epoch time
\State \textbf{continue}
\EndIf
\If{$ \frac{r_u}{d_u} < \bar{r} $}
\State $Q\text{.push}(u)$  // Postpone current active node to next
\State \textbf{continue}
\EndIf
\State $x_u = x_u + \alpha \cdot r_u$
\For{$v \in \nei(u)$}
\State $r_v = r_v + \frac{(1-\alpha) r_u}{d_u}$
\If{$r_v \geq \eps d_v$ and $v \notin Q$}
\State $\mc{Q}\text{.push}(v)$
\EndIf
\EndFor
\State $r_u = 0$
\State $t' = t' + 1$
\EndWhile
\State \textbf{Return} $\bm x$
\end{algorithmic}
\label{algo:forward-push-mean}
\end{algorithm}

\end{document}